\documentclass[aps,prx,reprint,preprintnumbers,superscriptaddress,nofootinbib,longbibliography,floatfix]{revtex4-2}
\usepackage{rotating}
\usepackage{array}
\usepackage{amsmath}
\usepackage[normalem]{ulem}
\usepackage{booktabs}
\usepackage[pdftex,table]{xcolor}
\usepackage{mathtools}
\usepackage{empheq}
\usepackage{multirow}
\usepackage{amssymb}
\usepackage{url}
\usepackage{physics}
\usepackage{xspace}
\usepackage{makecell}
\usepackage{nameref}
\usepackage{hyperref}

\hypersetup{
  colorlinks=true,
  citecolor=blue,
  linkcolor=blue,
  urlcolor=blue
}

\newcommand{\ttbar}{\ensuremath{\mathrm{t}\bar{\mathrm{t}}}\,}
\newcommand{\cathode}{\textsc{CATHODE}}

\begin{document}

\preprint{FERMILAB-PUB-25-0247-PPD}

\title{Weakly supervised anomaly detection with event-level variables}

\author{Liam Brennan}
\email{Liam.Robert.Brennan@cern.ch}
\affiliation{University of California Santa Barbara, Santa Barbara, CA, USA}
\author{Tamas Almos Vami}
\email{Tamas.Almos.Vami@cern.ch}
\affiliation{University of California Santa Barbara, Santa Barbara, CA, USA}
\author{Oz Amram}
\email{Oz.Amram@cern.ch}
\affiliation{Fermilab, Batavia, IL, USA}
\author{Sanjana Sekhar}
\affiliation{Johns Hopkins University, Baltimore, MD, USA}
\author{Yuta Takahashi}
\affiliation{University of Florida, Gainesville, FL, USA}
\author{Louis Moureaux}
\author{Manuel Sommerhalder}
\affiliation{University of Hamburg, Hamburg, Germany}
\author{Petar Maksimovic}
\affiliation{Johns Hopkins University, Baltimore, MD, USA}
\author{Tianji Cai}
\affiliation{School of Physical Science and Engineering, Tongji University, Shanghai, China}
\affiliation{State Key Laboratory of Autonomous Intelligent Unmanned Systems, MOE Frontiers Science Center for Intelligent Autonomous Systems, Tongji University, Shanghai, China}
\author{Nathaniel Craig}
\affiliation{University of California Santa Barbara, Santa Barbara, CA, USA}
\affiliation{Kavli Institute for Theoretical Physics, Santa Barbara, CA, USA}

\begin{abstract}
We introduce a new topology for weakly supervised anomaly detection searches, di-object plus~X.
In this topology, one looks for a resonance decaying to two standard model particles produced in association with other anomalous event activity (X). 
This additional activity is used for classification.
We demonstrate how anomaly detection techniques which have been developed for di-jet searches focusing on jet substructure anomalies can be applied to event-level anomaly detection in this topology. 
To robustly capture event-level features of multi-particle kinematics, we employ new physically motivated variables derived from the geometric structure of a collision's phase space manifold. 
As a proof of concept, we explore the application of this approach to several benchmark signals in the di-$\tau$ and di-$\mu$ plus~X final states. 
We demonstrate that our anomaly detection approach can reach discovery-level significances for signals that would be missed in a conventional bump-hunt approach. 
\end{abstract}

\maketitle
\flushbottom

\section{Introduction}
Despite many searches for new particles at the Large Hadron Collider (LHC), no clear evidence of beyond standard model physics (BSM) has been found. 
However, the space of possible signals which could be hiding in LHC data far exceeds the coverage of the dedicated searches performed thus far. 
In recent years, anomaly detection has been proposed as a new paradigm for conducting searches at the LHC~\cite{Kasieczka:2021xcg, Aarrestad:2021oeb, Belis:2023mqs}. 
Rather than designing a search to target a specific signal model, anomaly detection analyses take a model-agnostic approach. 
They use novel machine-learning-based strategies to identify events which differ from typical backgrounds without reference to a specific signal.
Both the ATLAS and CMS collaborations have now released searches involving these new anomaly detection approaches~\cite{ATLAS:2020iwa, ATLAS:2023azi, ATLAS:2023ixc,CMS:2024nsz, ATLAS:2025obc} and additional studies have been performed using CMS open data~\cite{Knapp:2020dde,Gambhir:2025afb}.

The majority of proposed anomaly detection algorithms can generally be split into two separate categories: unsupervised and weakly supervised algorithms. 
Unsupervised algorithms, first proposed in Refs.~\cite{Farina:2018fyg,Heimel:2018mkt}, are trained solely on background events and are then used to define an anomaly score which quantifies how similar an input event is to typical background events. 
Weakly supervised methods build on the classification without labels paradigm~\cite{Metodiev:2017vrx} to learn to distinguish between signal and background. 
This is accomplished by training a classifier to distinguish events in the signal region, which may contain some small fraction of signal, from a carefully constructed sample of background-only events~\cite{Collins:2018epr,Collins:2019jip}. 
Weakly supervised methods have the advantage that they are asymptotically optimal: the optimal weakly supervised classifier will learn the likelihood ratio, which is the ideal signal versus background classifier~\cite{Metodiev:2017vrx}.

The training procedure of weakly supervised anomaly detection methods requires a high quality estimate of the background in the high-dimensional feature space, allowing the classifier to identify the anomalous data events in the signal region which are distinct from background.
Defects in the background estimate will cause the classifier to learn the difference between the true background events in the signal region and the estimate, rather than identifying the signal.
This requirement means that simulations, which are known to have minor mismodelings, are generally not suitable for such a background estimate and data-driven methods must be used. 
Most weakly supervised anomaly detection methods have therefore focused on resonant signals, so the background composition in a signal mass window can be learned through an interpolation from sideband regions. Several approaches have been proposed to achieve the necessary data-driven high-quality background estimate~\cite{Collins:2018epr,Collins:2019jip, Amram:2020ykb, Andreassen:2020nkr, Hallin:2021wme, Raine:2022hht, Golling:2022nkl, Hallin:2022eoq, Sengupta:2023xqy, Das:2023bcj}. 
Many of these first anomaly detection approaches have focused on heavy resonances decaying to two boosted large radius jets, where well-studied jet substructure observables, with minimal correlations with the resonance mass, can be used as the classification features. 
There is also a large space of signal models which could produce many phenomenological varieties of jets, making this topology ideal for anomaly detection.
The weakly supervised searches from at the LHC \cite{ATLAS:2020iwa,CMS:2024nsz,ATLAS:2025obc} have also targeted this dijet topology.
Now that these first searches have been performed, it is timely to explore other signal topologies which are amenable to a weakly supervised resonant anomaly detection approach.

In this paper, we explore the application of weakly supervised anomaly detection in a new di-object~+~X topology using the \cathode\ methodology~\cite{Hallin:2021wme} with event-level features.
The \cathode\ method follows the general strategy described earlier and leverages a generative model trained on sideband data to characterize background distributions, subsequently interpolating these learned representations into the signal region to produce realistic background samples for weakly supervised anomaly detection. We consider signals which produce a two-body resonance in addition to other particles (X). 
Such a signature can easily arise if the resonance is produced through an extended decay chain or from associated production with additional particles.
These production mechanisms can lead to striking experimental signatures, with many objects in the event, potentially with rich kinematic correlations, and/or high missing energy. 
These distinct signatures can be leveraged to significantly reduce backgrounds; however, given the large phenomenological space of possibilities it is infeasible to design analysis strategies for every possible scenario. 
These two factors make for an ideal application of anomaly detection methods. 
This work develops a general framework for weakly supervised searches in this topology using event-level classification.
The usage of weakly supervised, rather than unsupervised, anomaly detection may also offer larger discovery potential for anomalies due to improved classification performance~\cite{Collins:2021nxn}.

As an exploratory study, we examine the specific case where the resonance decays into charged lepton pairs ($\mu$ or $\tau$). Since we expect similar behavior between electrons and muons, we focus our analysis on $\mu$ pairs instead of electrons.
Searches featuring $\tau$ leptons are strongly motivated by various BSM theories, which often preferentially couple to third-generation fermions, 
but are less explored at the LHC due to their more complex experimental signatures. 
Examples of searches include $Z^{'}$ bosons~\cite{ATLAS:2025oiy}, leptoquarks~\cite{CMS:2023qdw,ATLAS:2023uox,ATLAS:2021jyv}, and additional Higgs bosons~\cite{ATLAS:2020zms,CMS:2022goy}. 
These new particles are motivated by the observed anomalies in $B$-hadron decays to $D^{(\star)}\tau\nu$, which show enhanced rates compared to the corresponding decays involving lighter leptons~\cite{Capdevila:2023yhq,HFLAV:2022esi,LHCb:2015gmp,BaBar:2013mob,Belle:2015qfa,Belle:2017ilt,Belle:2019rba,LHCb:2017rln,LHCb:2023uiv}. 
If these anomalies are caused by heavy BSM particles, it is likely they also influence the di-$\tau$ production at high invariant masses.
Searches for $\mu$ lepton resonances are also strongly motivated by various BSM theories. While there are a plethora of experimental searches looking for BSM resonances decaying to second generation fermions~\cite{Cesarotti_2019,CMS_2307_08708, CMS_1803_06292, ATLAS:2017fih, ATLAS:2014pcp, Policicchio:2007rg,ATLAS:2018rjc}, it is likely there are still many unexplored exotic signatures.

Though we focus on the di-$\tau$ and di-$\mu$ final states in this work, the methodology we develop is entirely general and is applicable to other resonance~+~X searches,
such as di-electron, di-photon, or di-jet resonance~+~X. 
Although these other final states may have somewhat better existing experimental coverage, there is almost certainly still unexplored `X' signatures which could be covered by our approach. 
These other final states may also offer the advantage of larger samples available for the training of the background model of the \cathode\ method which can improve performance. 

Several other works have studied the application of anomaly detection methods to topologies beyond di-jet resonances.
In Refs.~\cite{Finke:2022lsu, Kasieczka:2024lxf} weak supervision was applied to topologies with high missing energy and a boosted jet while in Refs.~\cite{Bickendorf:2023nej,Curtin:2025ksm} the application of \cathode\ to supersymmetry scenarios producing pairs of gluinos was examined.
Recently, Ref.~\cite{Gambhir:2025afb} applied the \cathode\ method to CMS open data to demonstrate the identification of $\Upsilon$ particles decaying to muon pairs. 
This study used only features related to the dimuon resonance for classification rather than focusing on the other event activity we exploit here
A previous ATLAS search~\cite{ATLAS:2023ixc} applied an event-level unsupervised autoencoder-based anomaly detection algorithm to search for resonances for several di-object combinations, where at least one of the objects was required to be a jet.
These analyses used specific characteristics of these signal topologies to design the data-driven background estimate necessary for weak supervision, making generalization to other topologies unclear.
The resonance~+~X topology considered in this paper complements these prior works by offering a framework to apply weakly supervised anomaly detection to searches in many different final states.

The simulated data used in this paper can be found under Refs.~\cite{zenedo,zenedo3,zenedo2} and the analysis code is available on GitHub~\cite{github}.

\section{Setup}
\label{sec:setup}

\subsection{Final States}
Unlike di-jet searches where the QCD multijet background is dominant, di-$\tau$ and di-$\mu$ searches receive considerable backgrounds from multiple physics processes.
In the di-$\tau$ study, we focus on the final state where both $\tau$ leptons decay hadronically, which is the decay mode with the largest branching ratio. 
The dominant backgrounds in the all hadronic di-$\tau$ and the di-$\mu$ final states are Drell--Yan and \ttbar.
QCD and W+jets also contribute in this channel, with the fraction dependent on the purity of the $\tau$ and $\mu$ identification criteria being employed and pre-selection requirements. 
As this is an exploratory work, and these effects are not modeled well with simulation, we focus here on the Drell--Yan and \ttbar backgrounds only.
The methodology we employ should generally account for any background which varies smoothly in di-$\tau$ or di-$\mu$ mass. 
Due to the multiple decay modes of the $\tau$ and limited reconstruction efficiencies, di-$\tau$ searches feature significantly smaller event samples than the di-jet or di-$\mu$ searches of prior weakly supervised searches.
This presents an interesting challenge for the application of weakly supervised methods which require a large sample of background events in order to properly learn the probability distribution of the background events.

\subsection{The \cathode\ Method}
We apply the \cathode\ method for resonant anomaly detection~\cite{Hallin:2021wme}.
In \cathode, one defines a resonant variable, in our case the visible mass of the di-$\tau$ candidate ($m_\mathrm{vis}$) or invariant mass of the di-$\mu$ candidate ($m_{\mu\mu}$), and a set of features to be used for classification,~$\vec{x}$.
Signals are assumed to manifest in a localized region of the resonant variable and backgrounds are assumed to vary smoothly as function of the resonant variable.
These assumptions allow the probability distribution of the background to be learned from neighboring sideband regions.
The method proceeds in two steps.
First, a generative model is trained on the sideband regions to learn the probability density of $\vec{x}$ conditioned on $m$ for the background, $P_b(\vec{x} | m)$.
The trained model is then interpolated into the signal region and used to generate synthetic background events. 
A weakly supervised classifier is then trained to distinguish between the real data events $p_\mathrm{data}(\vec{x})$ in the signal region and the synthetic background events $p_\mathrm{bkg}(\vec{x})$. 
If there is a true signal in the signal region abundant in sufficient quantity, then the classifier will learn to identify its events, which will look anomalous with respect to the interpolated background.

The application of weak supervision to event-level anomaly detection raises interesting choices in feature selection.
Weak supervision is typically performed with a moderately-sized set of high-level features, and its performance is known to deteriorate with the use of many uninformative features~\cite{Finke:2023ltw}.
Due to the noise inherent to the weakly supervised training objective, the use of lower-level, higher-dimensional representations of the event can struggle to achieve good classification performance for moderate strength signals~\cite{Buhmann:2023acn}. 
Though this degradation may be mitigated in the classification step through the use of boosted decision trees as opposed to neural networks~\cite{Finke:2023ltw, Freytsis:2023cjr}, 
using a very high-dimensional feature set may cause difficulties in learning an accurate background estimate given the limited size of the training set available in final states such as di-$\tau$.

For these reasons, we explore the usage of the recently introduced~\cite{PS_theory,PS_pheno} phase space variables to encode the relevant kinematic features of the event in a compact representation.

\subsection{Phase Space Variables}

In Ref.~\cite{PS_theory, PS_pheno} a covariant description of the phase space of collider events involving a chosen number ($N$) of massless final-state particles was presented. 
The manifold of this $N$-body phase space, $\Pi_N$, was shown to be isomorphic to the product of an ($N-1$)-simplex ($\Delta_{N-1}$)
and a ($2N-3$)-hypersphere ($S_{2N-3}$), i.e. $\Pi_N \cong \Delta_{N-1}\times S_{2N-3}$.
A set of explicit global coordinates on the phase space manifold were introduced: 
$(N-1)$ coordinates $\rho$ on the simplex $\Delta_{N-1}$ 
and $(2N-3)$ angular coordinates $\xi, \chi$ on the hypersphere $S_{2N-3}$.
For $N=2$, the simplex (hypersphere) coordinates lie on a line (circle), while for $N=3$ they span a triangle (sphere).
Ref.~\cite{PS_pheno} used this representation to develop a metric between events produced at a hadron collider in terms of their phase space coordinates and demonstrated that the distance between events could be used effectively for event classification. 
Although the phase space coordinates are simply functions of the final-state particles' four-momenta, they are the natural variables for distinguishing different physics processes since they define the space on which $S$-matrix elements live.
In this work, we do not use the phase space metric itself, but rather directly use the phase space coordinates as an efficient representation of the kinematics of all the final state objects in the event.
We exploit the simplex coordinates only, since we found that they significantly enhance sensitivity while the generative model used for the background model, described in Section~\ref{sec:Methods}, has difficulties with the non-trivial topology of the hypersphere coordinates detailed in ~\ref{subsec:gen_model}.

At a hadron collider, in the final state's center of mass the simplex coordinates take the simple form
\begin{align}
\rho_i = \frac{1}{Q} (E_i \pm p_{z,i})
\end{align}
where $Q$ is the total energy of the event in the center of mass frame, the index $i = 1, \dots, N$ runs over the final-state particles, and the choice of sign is a matter of convention. The constraint $\sum_i \rho_i = 1$ implies that the $\rho_i$ are coordinates on an $N-1$ dimensional simplex. These simplex variables can be expressed in terms of rapidity $y$ and transverse momentum $p_T$ as
\begin{align}
\rho_i = \frac{p_{T,i}}{Q} e^{\pm y_i}
\end{align}
whereas the azimuthal angles of final-state particles are contained entirely in the sphere coordinates; we refer to Ref.~\cite{PS_pheno} for further details. The simplex coordinates are a useful compact combination of $p_T$ and $y$ insofar as the simplex contribution to the phase space distance between events is just the Euclidean distance between the simplex coordinates. Indeed, we find that the simplex coordinates enhance sensitivity compared to a feature set of the same dimension using $p_T$; see Appendix~\ref{sec:pt_analysis} for a detailed study.

It should be noted that the phase space coordinates encode only the kinematics of the event, and therefore, do not capture all event features that may be useful for event-level anomaly detection. 
For example, they do not encode which type of object in the event (jet, lepton, photon, etc) is producing said kinematics.
It therefore may be useful to supplement phase space variables with additional variables which encode object-type information. 
The phase space variables are also normalized such that they depend only on the relative fraction of energies carried by each object and its orientation.
They do not contain information about the overall energy scale of the event; i.e. an event with 100\,GeV of energy split evenly among 10 jets in a certain topology is encoded the same as an event with 1\,TeV split among those jets in the same topology. 
We therefore supplement such scale information to our feature set used for classification.

\section{Simulated Samples}
\label{sec:samples}

We produced simulated samples with \verb|MadGraph_aMC@NLO|~\cite{Alwall:2014hca} v3.5.4 using UFOs~\cite{Darme:2023jdn} from the FeynRules database. We set the center of mass energy to 13~TeV in order to match the Run-2 conditions of the LHC. Hadronization was done with \verb|Pythia8|~\cite{Sjostrand:2014zea} v8.306. All samples are generated at leading order using the NNPDF2.3 LO parton distribution function~\cite{Ball:2010de, NNPDF:2014otw}, with default dynamic scale choices in \verb|MadGraph|.

For the SM background samples we used the SM UFO~\cite{feynrulesSM} with CKM mixing, and with the four-flavor scheme. We simulated extra jets (ISR/FSR) up to 2 jets, i.e. 0, 1, or 2 jets besides the hard scatter. Matrix element and parton shower matching was performed using the MLM scheme~\cite{Alwall:2007fs} with a matching scale of 30\,GeV. For the $t\bar{t}$ sample, the top quarks were decayed in \verb|MadSpin|~\cite{Artoisenet:2012st} to W bosons that decay to $\tau$ or $\mu$ leptons, denoted commonly as $\ell$. In order to enhance the generation efficiency, we applied a generator level cut on the di-$\ell$ invariant mass of 120\,GeV. The cross section of the Drell--Yan sample is $17.37$\,pb and the cross section of the $t\bar{t}$ sample is $13.49$\,pb. Sufficient events were generated to have an equivalent luminosity of 138\,fb$^{-1}$, to match that of the proton-proton collisions data taken during Run-2 (2016-18) of the LHC.

\begin{figure*}[!ht]
    \includegraphics[width=0.25\linewidth]{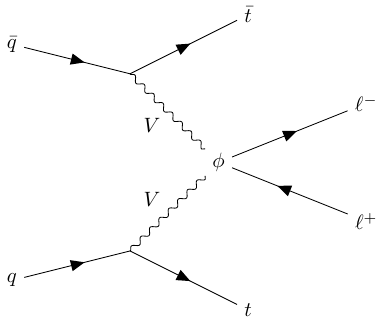}
    \hspace{20pt}
    \includegraphics[width=0.25\linewidth]{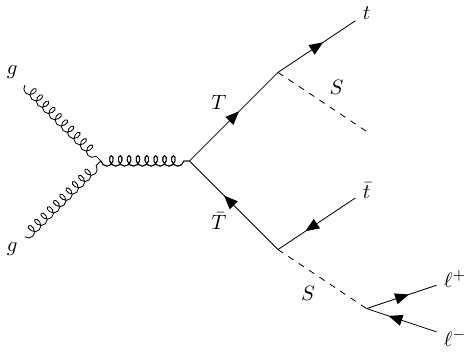}
    \hspace{20pt}
    \includegraphics[width=0.25\linewidth]{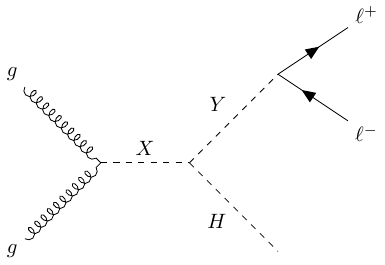}
    \\
    \caption{Feynman diagrams of the $t\bar{t}\phi$ (left), $T'\bar{T'}$ (middle), and $X \to YH$ (right) signal models. The $\ell$ symbol denotes the $\tau$ or $\mu$ leptons. The $S^0$ decays inclusively according to its branching ratios. The $H$ boson in the $X \to YH$ model decays invisibly. }
    \label{fig:feynman}
\end{figure*}

We explored a set of several BSM models as benchmarks to evaluate the performance of our approach.
The details of the generation are summarized in Table~\ref{tab:samples}, where the number of events is after the generator cuts are applied. Some example Feynman diagrams of the signal processes are shown in Fig.~\ref{fig:feynman}. For simplicity, we set the masses of the di-$\ell$ resonance in all models to 250\,GeV.

In the \verb|2HDM| model~\cite{feynrules2hdm,Branco:1999fs,Degrande:2014vpa} we use the heavy Higgs particle, denoted with $\Phi$, and let it decay to an $\ell$ lepton pair. In the \verb|eVLQ| model~\cite{feynrules2eVLQ,Banerjee:2022xmu}, we let the vector-like quark, $T'$ decay to a SM top and a BSM scalar $S^0$. One of the $S^0$ then decays to an $\ell$ lepton pair while the other decays inclusively according to branching ratios  similar to a heavy Higgs boson. In the \verb|NMSSM| model~\cite{feynrules2NMSSM,Ellwanger:2009dp,Djouadi:2008uw,Skands:2003cj,Allanach:2008qq} we simulate cascade processes, where the heavy resonance X decays to a Higgs boson (H) and an unknown particle Y. Y further decays to the $\ell$ leptons and the H decays invisibly, leading to missing transverse momentum. 
We explored non-resonant signals such as vector-like anti-leptons~\cite{Bigaran:2023ris}, leptoquarks~\cite{Baker:2019sli} and heavy neutral leptons~\cite{Kayser:1981nw}. 
We did not achieve significant improvements with our method on these signals, as expected since these models are not well localized to the signal region, making them difficult to identify with the \cathode\ approach.
We therefore did not study them further.

\begin{table}[!ht]
    \centering
     \setlength{\tabcolsep}{4pt}
    \scalebox{0.75} {
    \begin{tabular}{ccccc}
    \toprule[1.5pt]  
    UFO & Name & Extra jets & Parameters & \#Events  \\
    \midrule[\heavyrulewidth]
    SM & Drell--Yan $\ell \bar{\ell}$ & up to 2 & N/A & 3M  \\ 
    \cmidrule{1-5}  
    SM & $t\bar{t}\to W^-W^+b\bar{b}\to \ell \bar{\nu} + \bar{\ell} \nu$  & up to 2 & N/A & 2M \\ 
    \midrule[\heavyrulewidth]
    2HDM & $t\bar{t}\Phi\to\ell\bar{\ell}$ &  0 & $m_\Phi = 250$\,GeV & 100k \\ 
    \cmidrule{1-5}
    \multirow{2}{*}{eVLQ} & \multirow{2}{*}{$ T'\bar{T}'\to t S^0 \bar{t} S^0 \to \ell\bar{\ell}+\text{incl}$} & \multirow{2}{*}{0} & \multirow{2}{*}{\shortstack{$m_{T'}= 1000$\,GeV \\ $m_S = 250$\,GeV}} & \multirow{2}{*}{100k} \\ 
    & & & & \\
    \cmidrule{1-5}
    \multirow{3}{*}{NMSSM} & \multirow{3}{*}{$X \to YH \to \ell \bar{\ell} + \text{inv}$} & \multirow{3}{*}{0} & \multirow{3}{*}{\shortstack{$m_{X}= 800$\,GeV \\ $m_H = 400$\,GeV \\ $m_Y = 250$\,GeV}} & \multirow{3}{*}{100k} \\ 
    & & & & \\
    & & & & \\
    \bottomrule[1.5pt]  
    \end{tabular}
    }
    \caption{List of samples used, and the details of the production. The $\ell$ symbol denotes the $\tau$ or $\mu$ leptons.}
    \label{tab:samples}
\end{table}

We used \verb|Delphes|~\cite{deFavereau:2013fsa} to simulate detector and reconstruction effects, using the \verb|CMS| detector settings. Jets were clustered using the anti-$k_T$ algorithm with $R=0.4$ as implemented in \verb|FastJet|\cite{Cacciari:2011ma} via \verb|Delphes|. $\tau$ jets were reconstructed using the default CMS \verb|Delphes| $\tau$ tagging module with $p_T > 20$\,GeV and $|\eta| < 2.5$. A 1\% rate of mis-tagging QCD jets as $\tau$'s and a 60\% $\tau$ tagging efficiency were included~\cite{CMS:2018jrd,CMS:2022prd}. No dedicated trigger simulation was included.

\begin{figure*}[!ht]
    \includegraphics[width=0.47\linewidth]{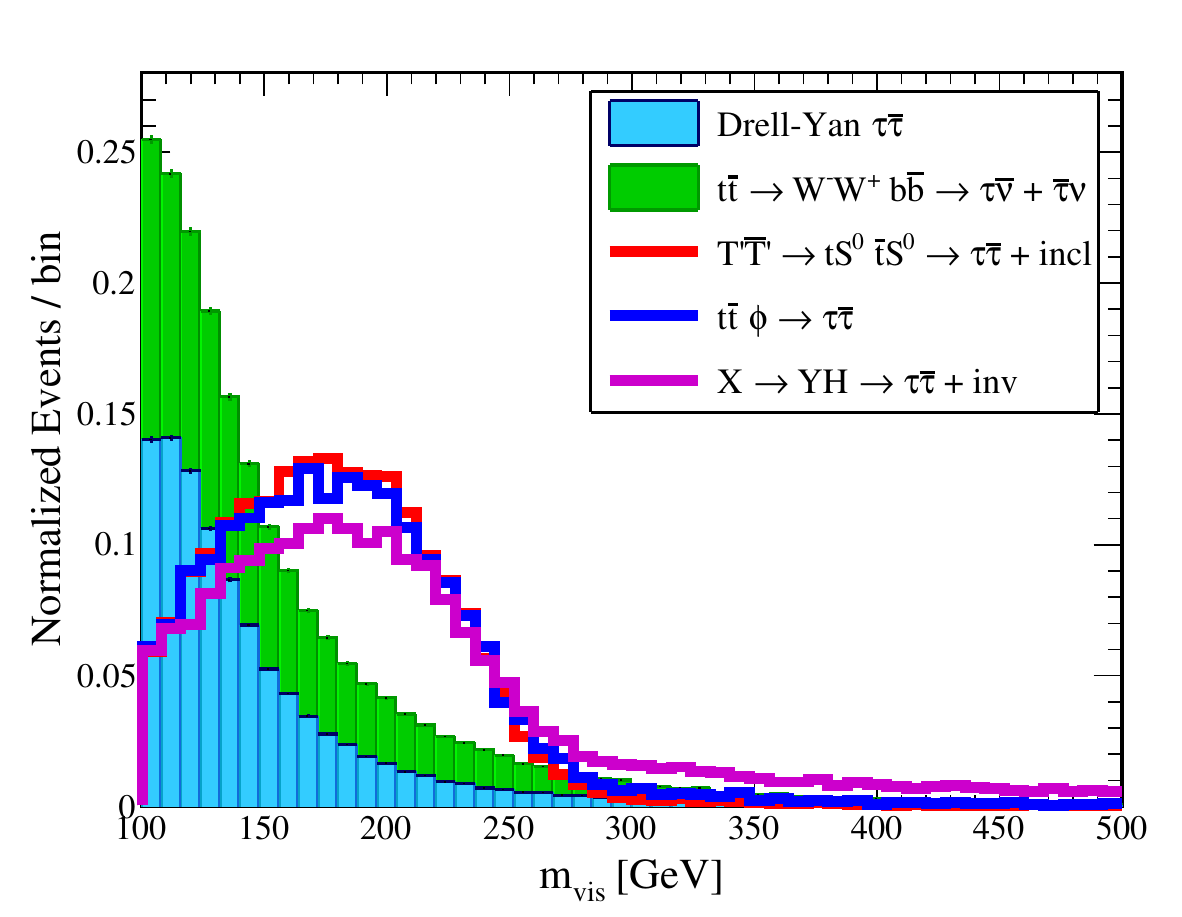}
    \includegraphics[width=0.47\linewidth]{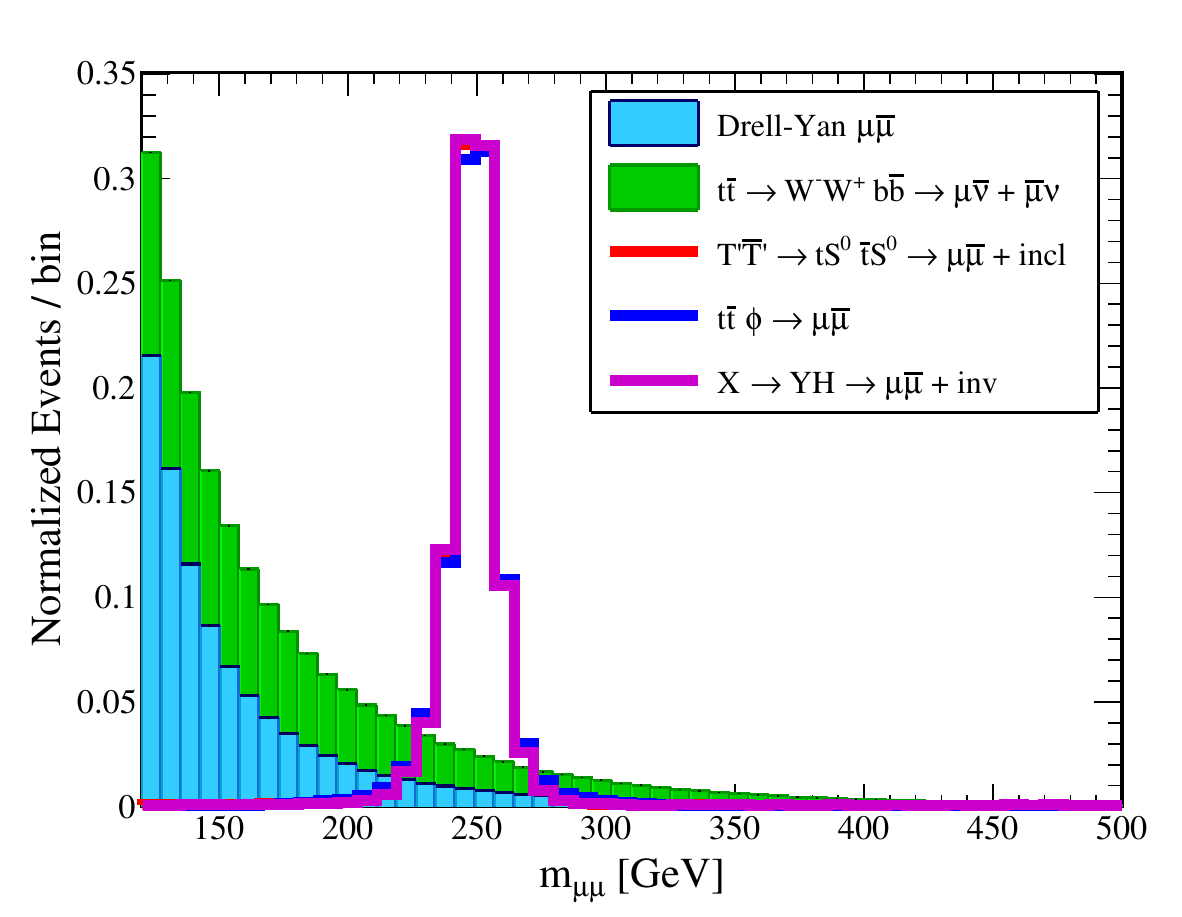}
    \caption{Invariant visible mass distribution of the di-$\tau$ system (left) and the di-$\mu$ system (right)}.
    \label{fig:inv_mass}
\end{figure*}

Figure~\ref{fig:inv_mass} shows the reconstructed $m_\mathrm{vis}$ and $m_{\mu\mu}$ distributions of the di-$\tau$ and di-$\mu$ systems respectively. While the generator level mass cut is at 120\,GeV, because some of the di-$\tau$ momentum is lost to the neutrinos, the reconstructed visible mass is smoothly falling starting from 100\,GeV. Histograms are normalized to unit area.

\section{Methods}
\label{sec:Methods}

For the di-$\tau$ channel, we select the two highest $p_T$ $\tau$-tagged jets with $\abs{\eta} < 2.4$ as our resonance candidate. 
The leading $\tau$ jet is required to have $p_{T} > 40$~GeV.
For the di-$\mu$ channel, we select the two highest $p_T$, oppositely charged, muons with $\abs{\eta} < 2.4$ and $ p_T> 20$~GeV.
We use the $m_\mathrm{vis}$ and $m_{\mu\mu}$ as the resonance variable in the \cathode\ approach in each channel.
We require $m_\mathrm{vis} > 100$~GeV ($m_{\mu\mu} > 120$~GeV) in the di-$\tau$ (di-$\mu$) channel.
For simplicity, we consider a SR with a mass window of 175--300\,GeV for di-$ \tau $ and 230--270 for di-$ \mu $, both of which contain the majority of our generated signals, as visible in Fig.~\ref{fig:inv_mass}.
In a real search, a scan could be performed repeating the procedure for different signal window choices to cover the full range as has been done in previous anomaly detection searches. We expect the analysis to remain effective because the phase space variables should continue to capture the kinematic differences between signal and background for arbitrary values of the mass scale.
When resonant SM backgrounds are present, their mass windows are masked so the flow learns the background density from uncontaminated sideband regions.

Within the SR, we reserve 15\% of background events as a test set for performance evaluation and an equal number of signal events. The remaining 85\% of SR background events were divided equally between training (42.5\%) and validation (42.5\%) sets.
In a true search, a cross-validation approach would be employed to ensure data events are not reused for training and signal extraction, as had been done in prior weakly supervised searches~\cite{ATLAS:2020iwa, CMS:2024nsz, ATLAS:2025obc}. 
Signals are injected into the SR with varying strengths to test performance, as discussed in Sec.~\ref{sec:results}.

\subsection{Features}

To establish a consistent representation of the phase space manifold, we require fixed dimensionality of objects across all events. 
We define events in terms of constituent objects (jets, electrons, muons, or photons), with each event represented by exactly fifteen objects. For events with fewer than fifteen physical objects, we perform zero-padding with four-vectors as $(E,p_x,p_y,p_z) = (0.1, 0.1, 0, 0)$. We avoid four-vectors with exactly zero momentum because this leads to division by zero in the phase space coordinate calculation, causing numerical instabilities.  Objects are ordered by decreasing transverse momentum to ensure consistent event representation. 

The fifteen objects yield fifteen simplex coordinates per event, following the manifold construction described in Section~\ref{sec:setup}.

Our analysis reveals that the first ten simplex coordinates, which corresponds to the ten highest $p_T$ objects,  capture the majority of the relevant phase space topology, with the remaining coordinates primarily corresponding to the zero-padded objects in typical events.

To complement the phase space variables, we incorporate event-level kinematic observables: the scalar sum of non-$\tau$ jet $p_T$, the missing transverse momentum (MET), and the angular separation ($\Delta$R  $=\sqrt{(\Delta\eta)^2+(\Delta\phi)^2}$, where $\eta$ is the pseudorapidity and $\phi$ is the azimuthal angle) between the two leptons. 
These features encode global event characteristics that are not fully captured by the phase space coordinates alone. 
We focus on these continuous kinematic features, rather than discrete quantities like jet multiplicity, because normalizing flows require differentiable and invertible transformations.
The scalar sum of jet $p_T$ encodes valuable information on jet kinematics and overall event structure, while preserving the continuity required for a normalizing flow.
Our final feature vector consists of: the first ten simplex coordinates, MET, non-$\tau$ jet $p_T$ sum, and lepton $\Delta$R.


\subsection{Classification and Background Evaluation}

Our generative model uses conditional flow matching~\cite{lipman2023flowmatchinggenerativemodeling,tong2024improvinggeneralizingflowbasedgenerative, holderrieth2025introductionflowmatchingdiffusion}, and is trained in the sidebands.
For simplicity, the training is performed on a background-only sideband sample, assuming the signal contamination to be negligible.
A kernel density estimate is used to learn the $m_\mathrm{vis}$ or $m_{\mu\mu}$ distribution for use in synthetic background generation.
Details of the model architecture and exact generation procedure are given in Appendix~\ref{subsec:gen_model}.

After training, we oversample events from the background model to generate synthetic background events in the SR at 3 times the amount of data in the SR. This oversampling provides the classifier with additional background examples to learn from, improving the robustness of the weakly supervised classification. Class balance is maintained through an appropriate weighting in the loss function. 

For the weakly supervised classification task, we employ an ensemble of 50 histogram-based gradient boosted decision trees (BDTs). 
Ensembles of BDTs have demonstrated superior performance over neural networks in weakly supervised training with high-dimensional feature spaces~\cite{Finke:2023ltw}. Our implementation utilizes the \texttt{HistGradientBoosting} classifier from \texttt{scikit-learn}~\cite{Scikit_learn:2011}. Hyperparameters for the BDT are kept to be the defaults as in the \verb|SK_CATHODE|~\cite{sk_cathode} repository.

Once we have trained the weakly-supervised classifier, we define our final event selection based on a cut on the classifier score that is 1\% efficient on the background.
Though tighter cuts are seen to yield higher signal enhancements, this threshold is seen to give good classification performance for all signals, and is realistic to thresholds used in previous weakly supervised anomaly detection searches~\cite{ATLAS:2020iwa, CMS:2024nsz, ATLAS:2025obc}. 

\subsection{Synthetic Background Quality Validation}

\begin{figure*}
    \centering
    \includegraphics[width=0.32\linewidth]{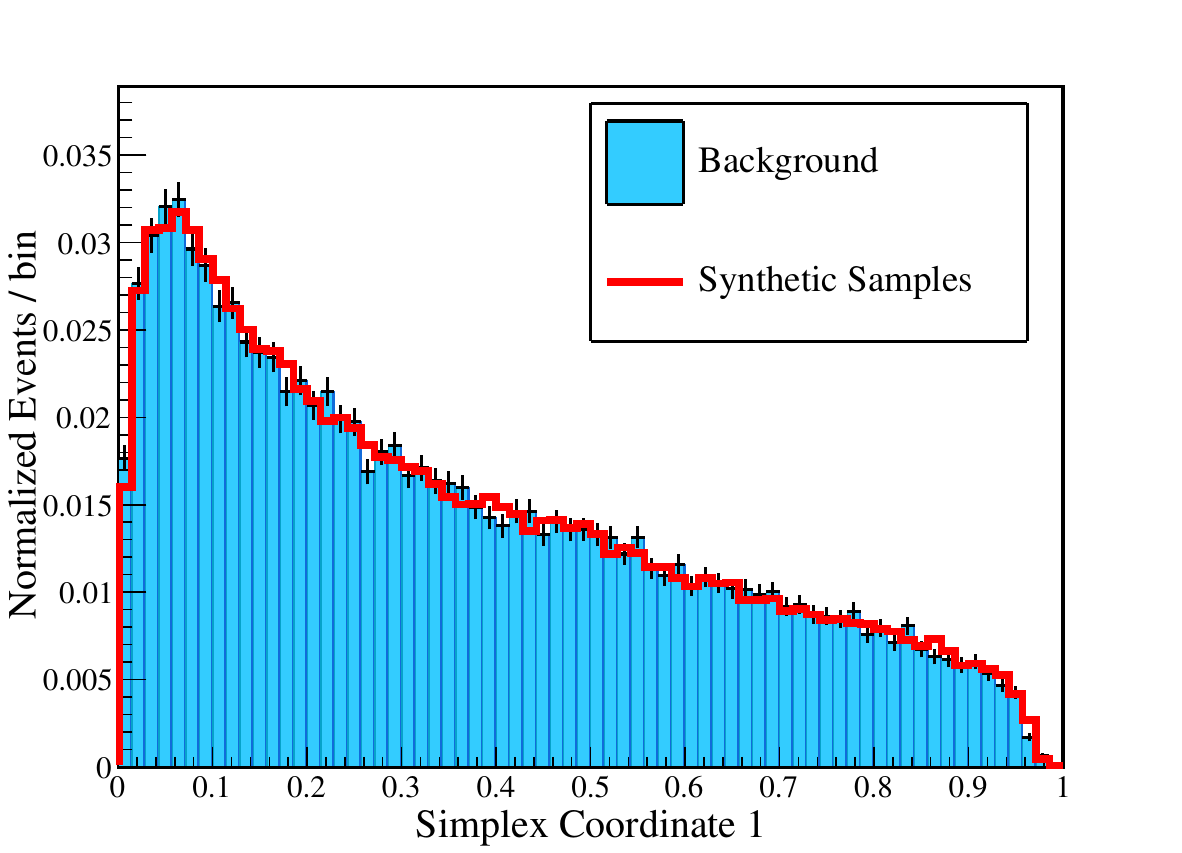}
    \includegraphics[width=0.32\linewidth]{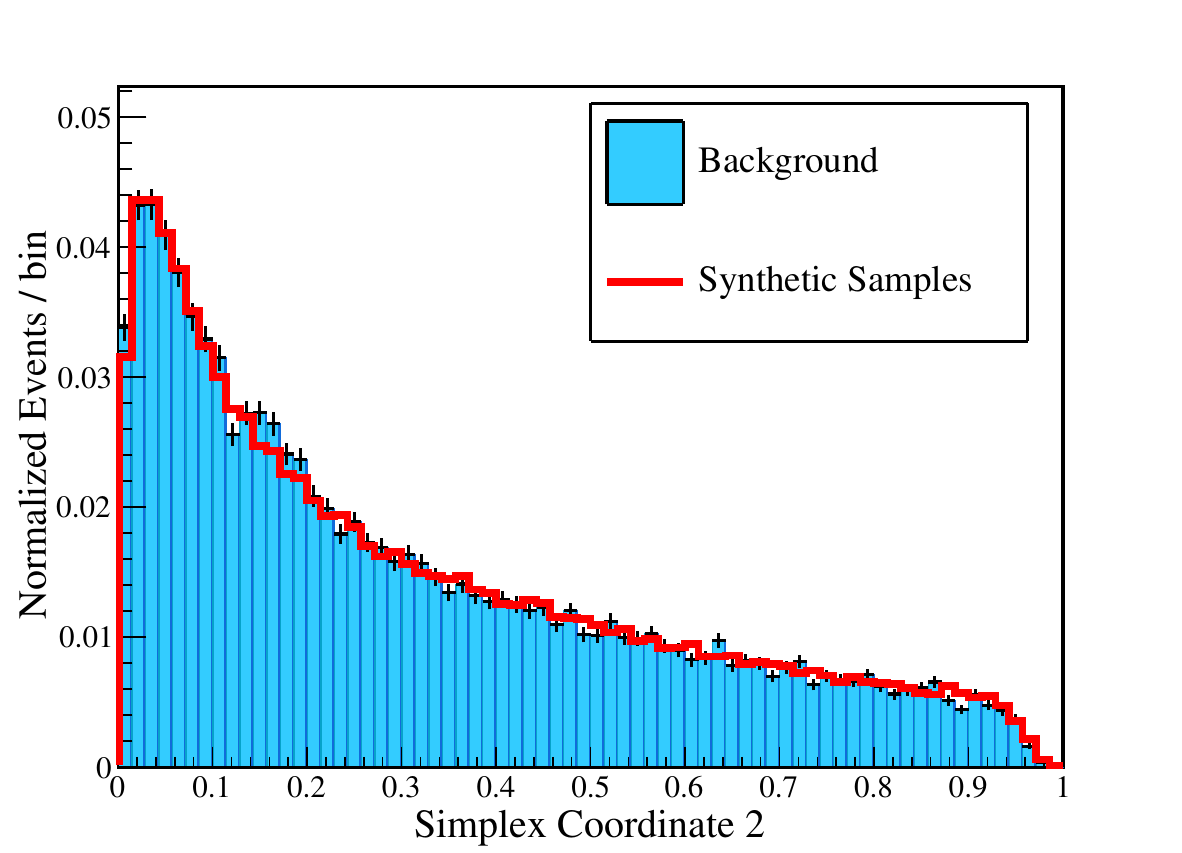}
    \includegraphics[width=0.32\linewidth]{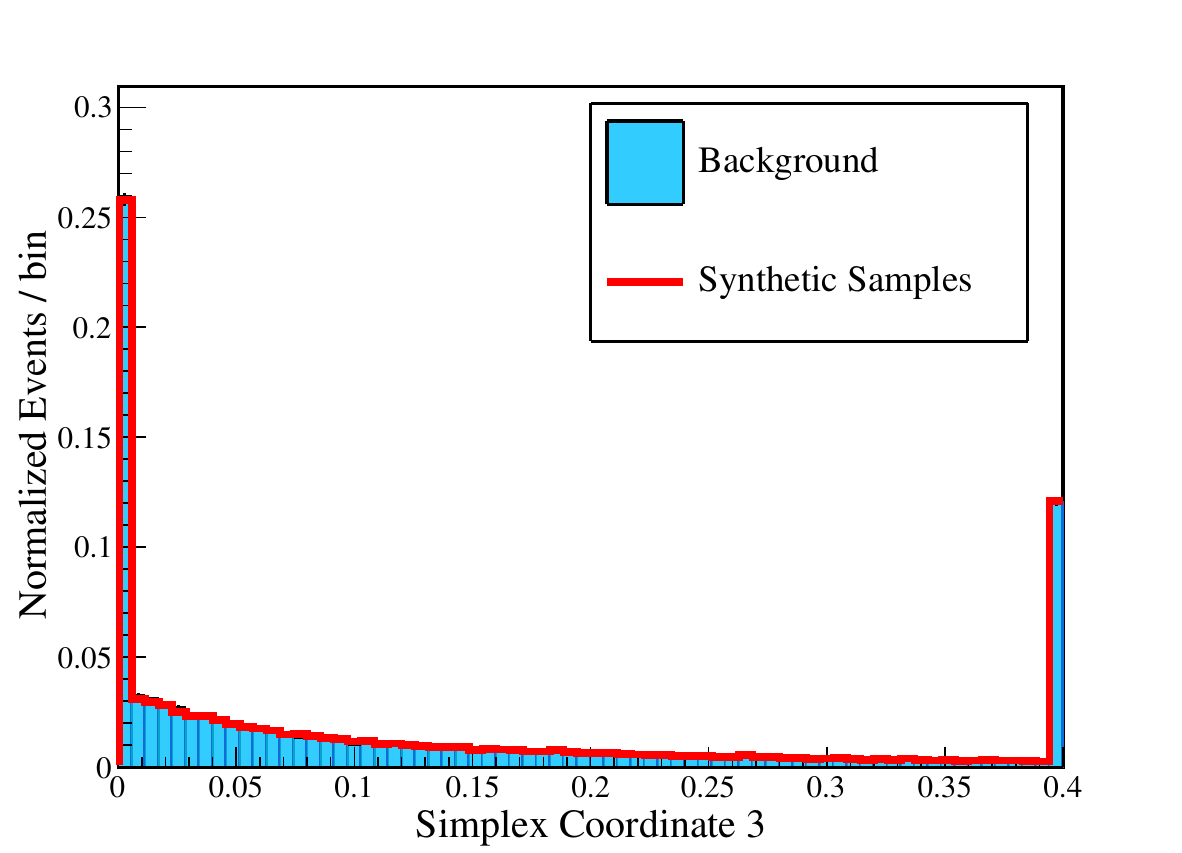} \\
    \includegraphics[width=0.32\linewidth]{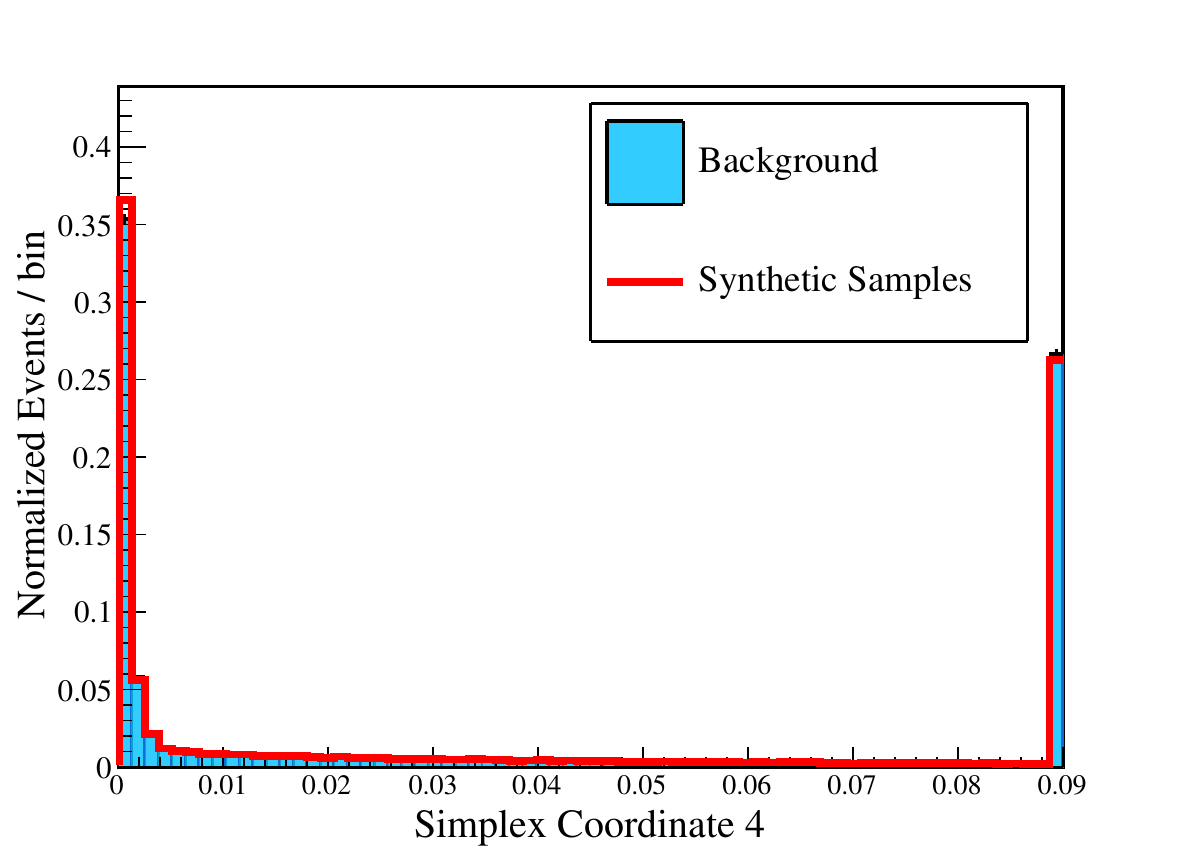}
    \includegraphics[width=0.32\linewidth]{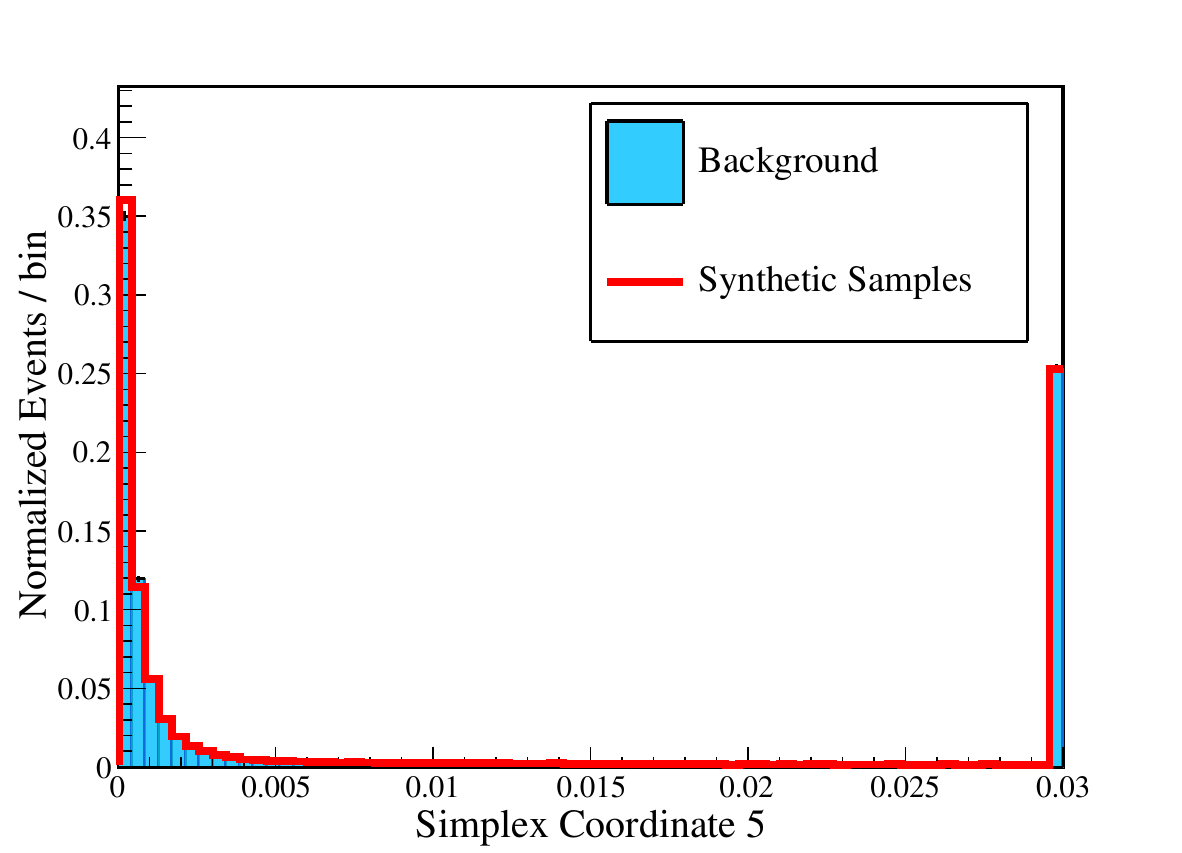}
    \includegraphics[width=0.32\linewidth]{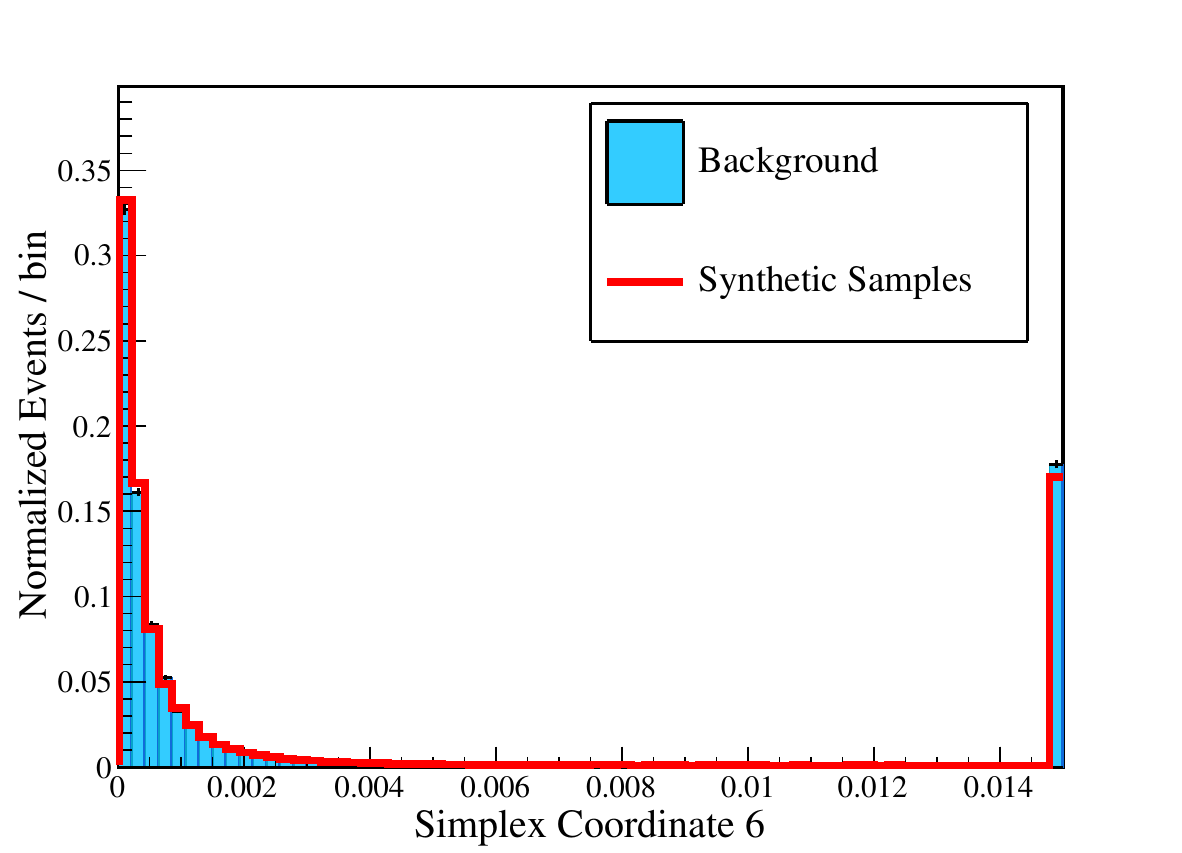} \\
    \includegraphics[width=0.32\linewidth]{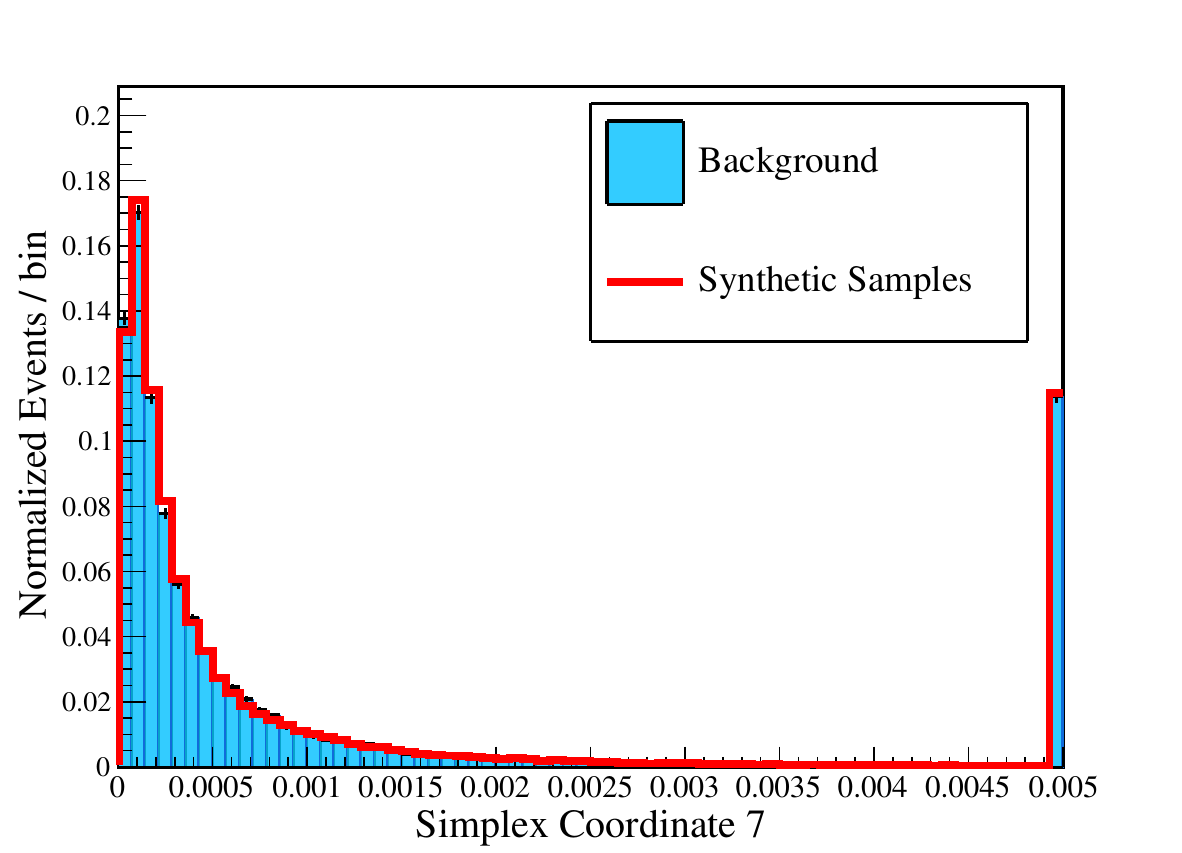}
    \includegraphics[width=0.32\linewidth]{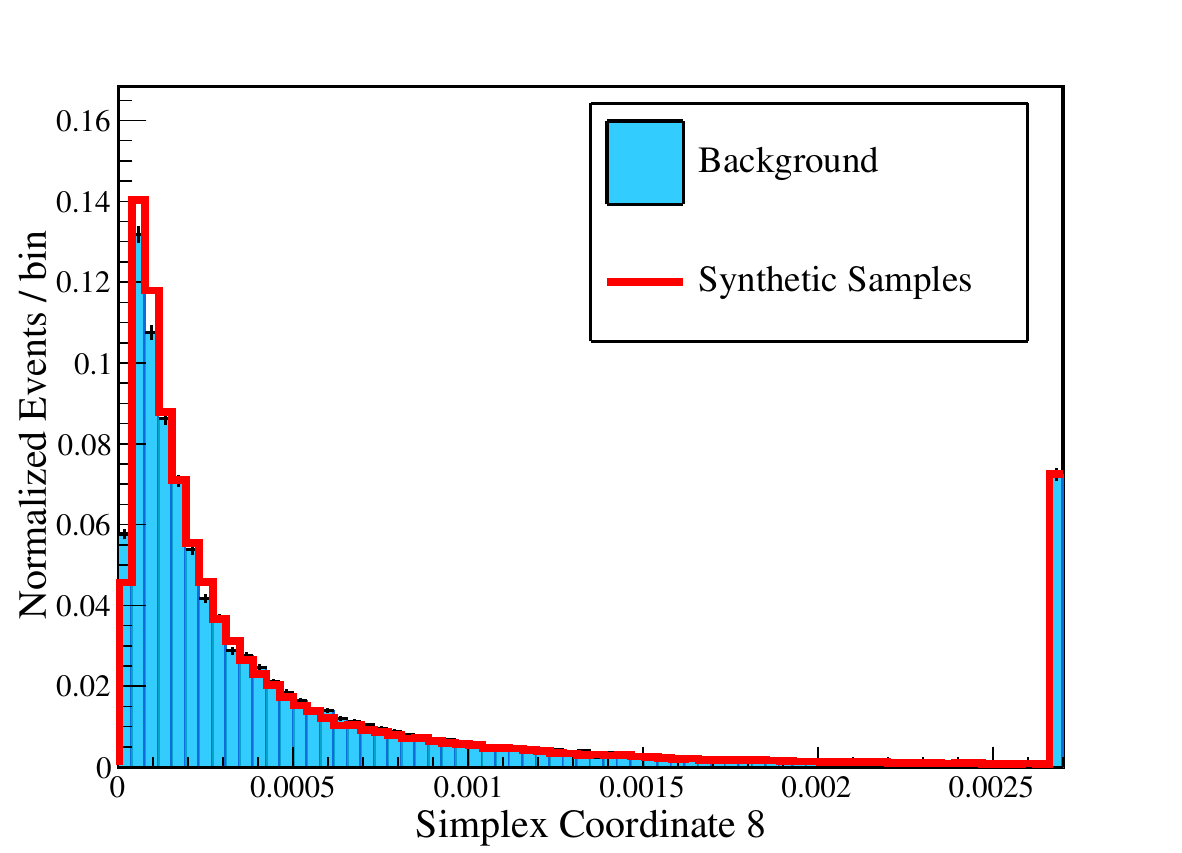}
    \includegraphics[width=0.32\linewidth]{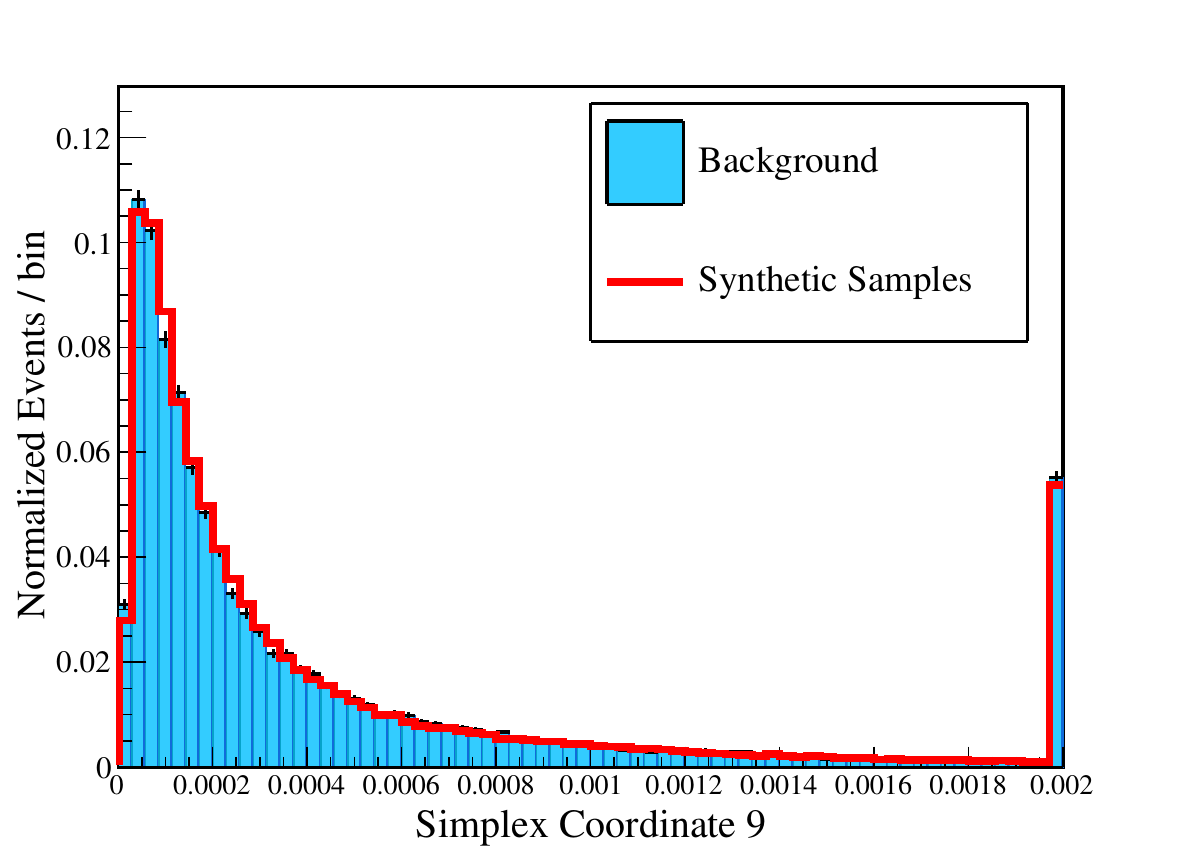} 
    \includegraphics[width=0.32\linewidth]{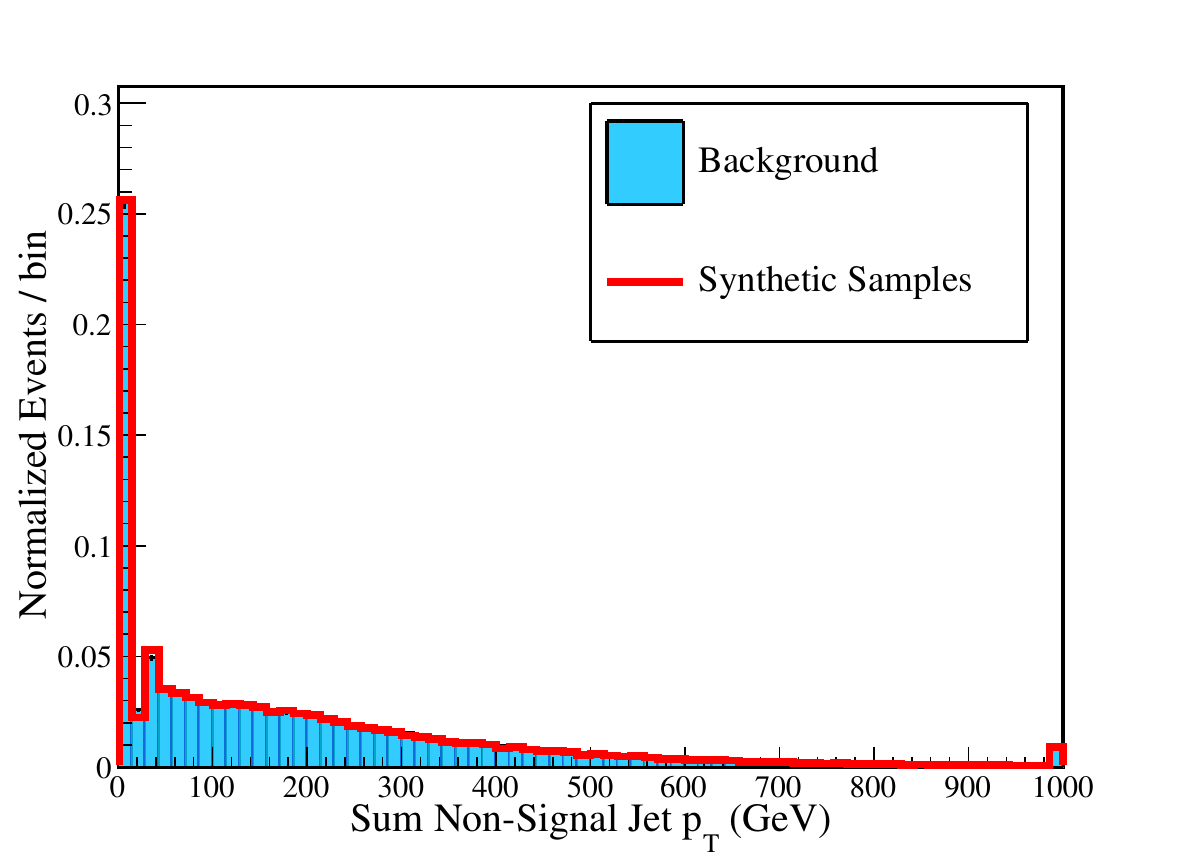}
    \includegraphics[width=0.32\linewidth]{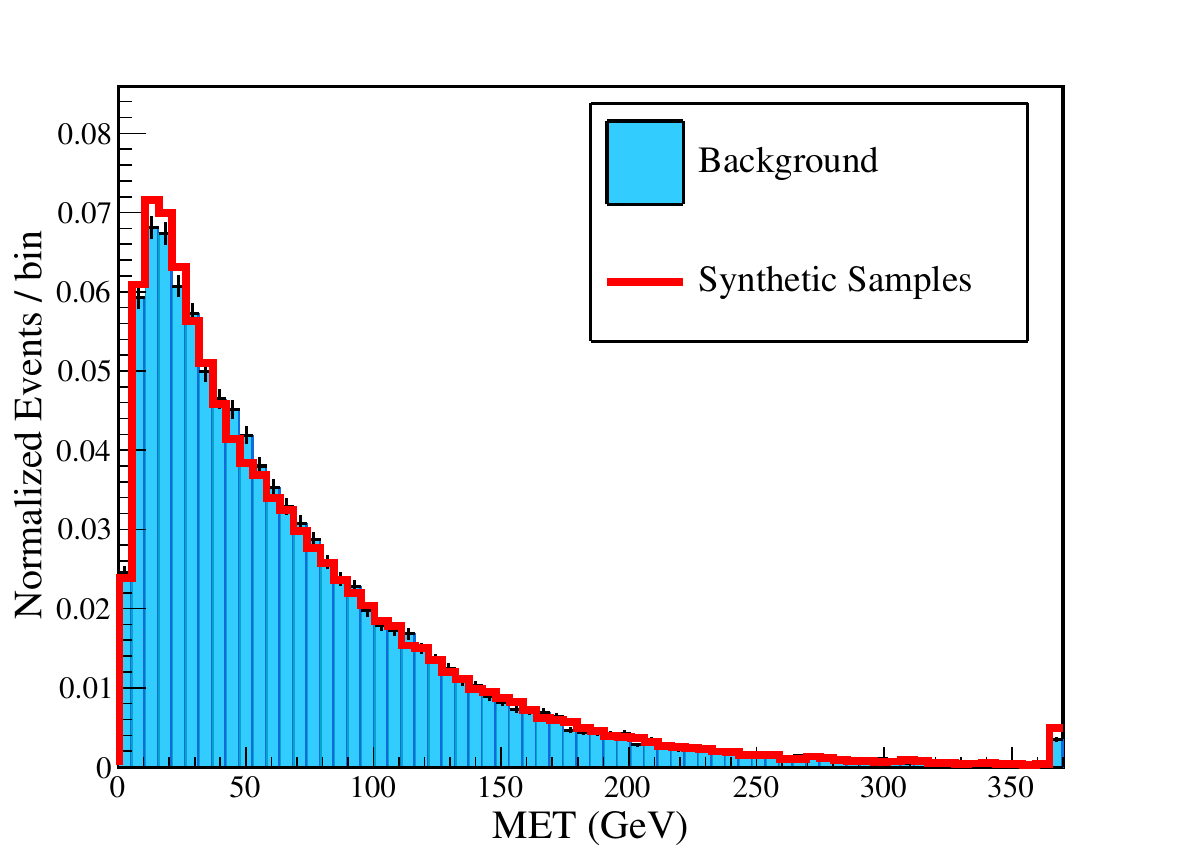} 
    \includegraphics[width=0.32\linewidth]{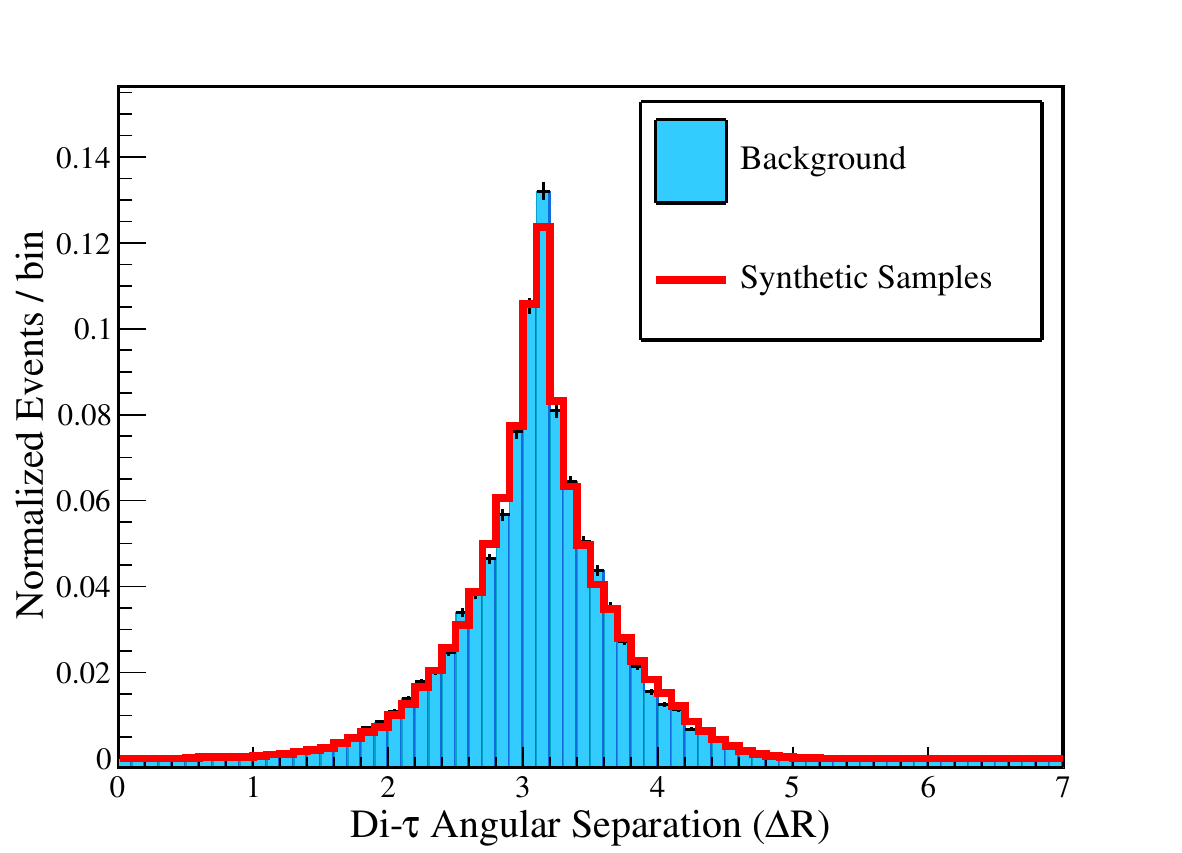} 

    \caption{A comparison of the distributions of true background events in the signal region and the synthetic ones generated by our normalizing flow for di-$\mu$. Simplex coordinate 10 is omitted for brevity, but has similar distributions.The last bin in each histogram shows overflow entries beyond the range of the histogram axis. The flow is seen to model most of the distributions well. Some minor discrepancies are seen in the modeling close to variable boundaries at zero. The feature distributions and agreement between synthetic and true are similar in the di-$\tau$ channel.}
    \label{fig:feats}
\end{figure*}

We mix two well-performing generated backgrounds in equal proportions, as each may capture different aspects of the background distribution. A comparison of the synthetic events generated by the background model and the true background events in the SR is shown in Fig.~\ref{fig:feats}.

To validate the quality of our synthetic background, we applied our BDT classifier to distinguish real background events from synthetic ones in the signal region. The BDT classifier uses same architecture and features as are used in the weakly supervised training.
The training corresponds to what the weakly supervised classifier would see if there were only background events in the signal region. 
A near-random classification performance shows that such a classifier cannot tell the samples apart. 
This indicates that if there was instead signal events in the signal region the classifier would focus on correctly classifying them to improve the loss, rather than concentrating on the differences between the synthetic and true background. This test yields a classification area under curve of 0.52 for both channels, close to the random-chance value of 0.5.

Imperfect modeling of the background may dampen sensitivity in the \cathode\ approach, as the weakly-supervised classifier may focus on the differences between the synthetic and true backgrounds rather than identifying signals.
However, even with our imperfect background modeling, we have seen strong anomaly detection performance with our chosen signals.
We therefore conclude our background model is of sufficiently high quality for the anomaly detection procedure to work, and leave further improvements to future studies.

\section{Results}
\label{sec:results}

\begin{figure*}
    \centering
    \includegraphics[width=0.49\linewidth]{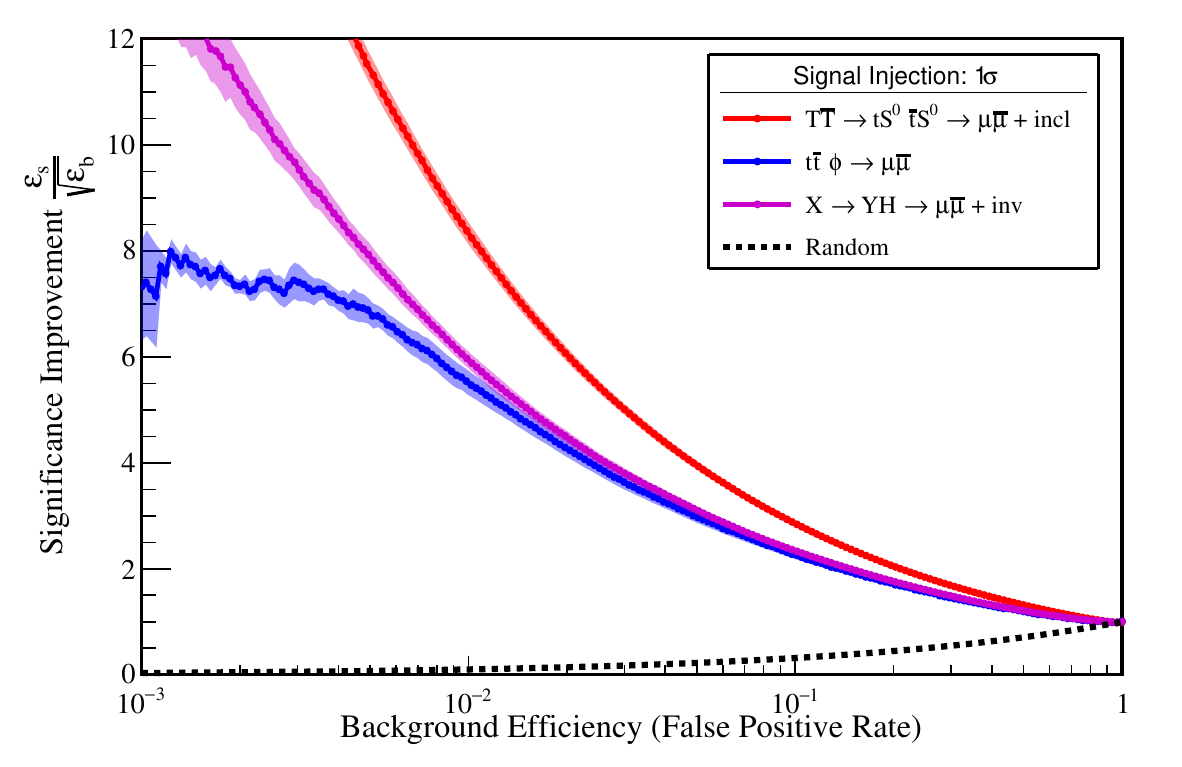}
    \includegraphics[width=0.49\linewidth]{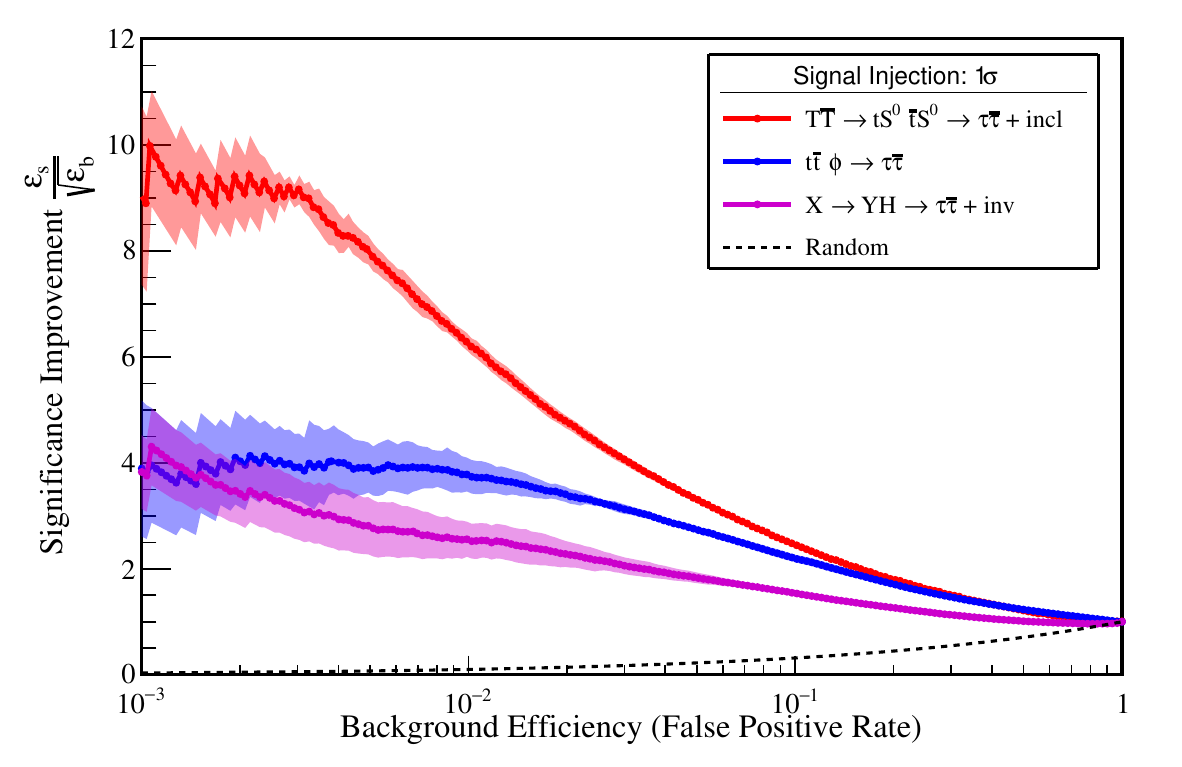}
    \includegraphics[width=0.49\linewidth]{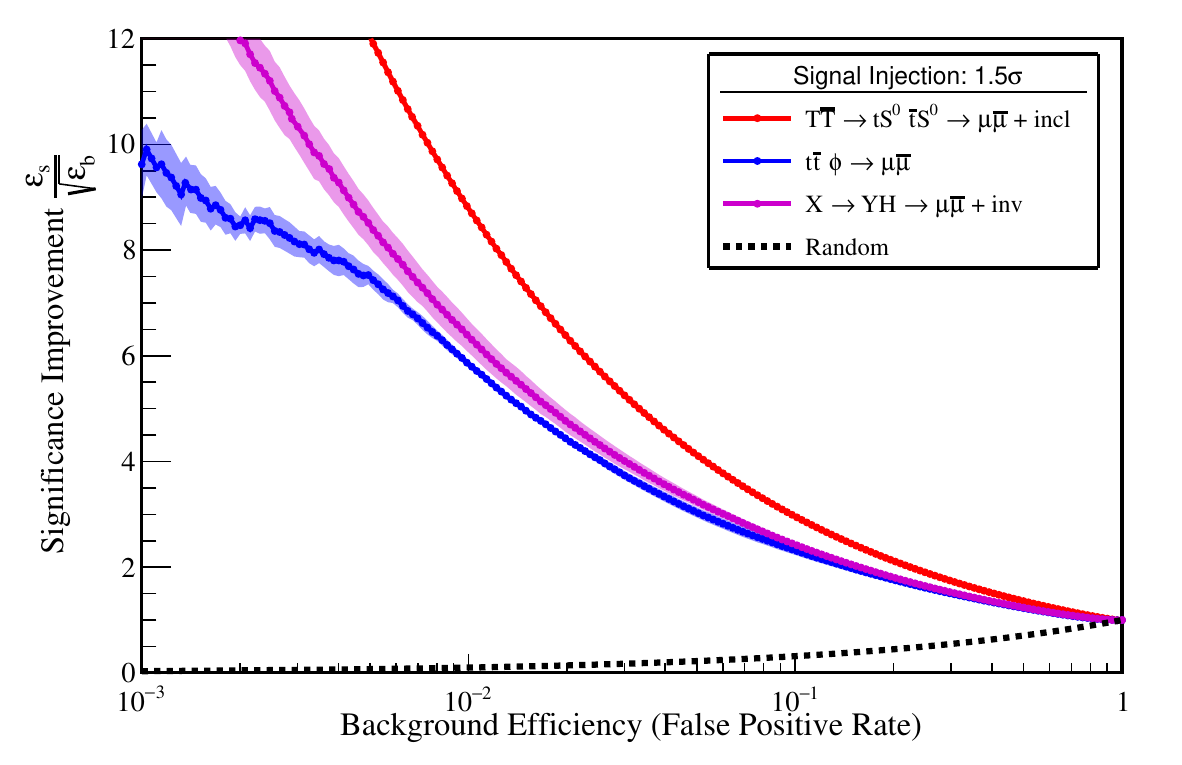}
    \includegraphics[width=0.49\linewidth]{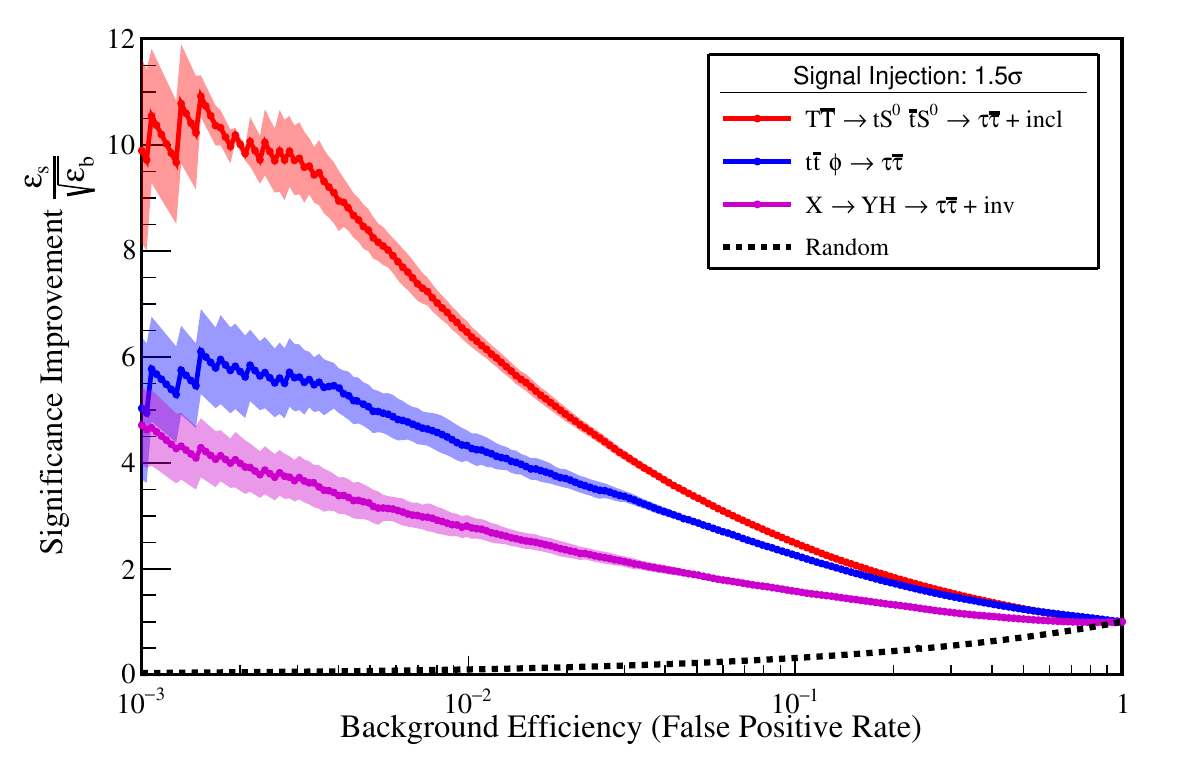} \\
    \caption{Significance improvement curves showing classification performance on across our different signal models for both di-$\mu$(left) and di-$\tau$(right)  for signal injections of 1$\sigma$ (top), and 1.5$\sigma$ (bottom). The shaded bands represent the statistical uncertainty (±1 standard deviation) in the measured average significance improvement from the 5 different runs.}
    \label{fig:sics}
\end{figure*}

The performance of the signal classification is expected to vary as a function of the amount of signal present in the dataset. 
We run the training procedure for several different injection strengths for each signal and evaluate the resulting classification performance.
For each injection the procedure is repeated 5 times to assess the stability of results.
The injection strength is quantified by the approximate signal significance, $\sigma = \frac{S}{\sqrt{B}}$, prior to any additional selection from the anomaly detector. 
We test injection strengths of 0.25$\sigma$, 0.5$\sigma$, 1$\sigma$, 1.5$\sigma$, and 2$\sigma$. 
Significance improvement curves (SIC) are used to quantify classification performance. The curves are created by scanning across different classifier thresholds and plotting the signal efficiency divided by the square root of the background efficiency $\frac{\epsilon_s}{\sqrt{{\epsilon_b}}}$, as a function of the background efficiency.
The SIC shows the approximate enhancement in signal statistical significance which can be achieved by cutting at a given background efficiency.
The achieved classification performance for different signal injections is shown in Fig.~\ref{fig:sics}.

\begin{figure*}
    \centering
    \includegraphics[width=0.44\linewidth]{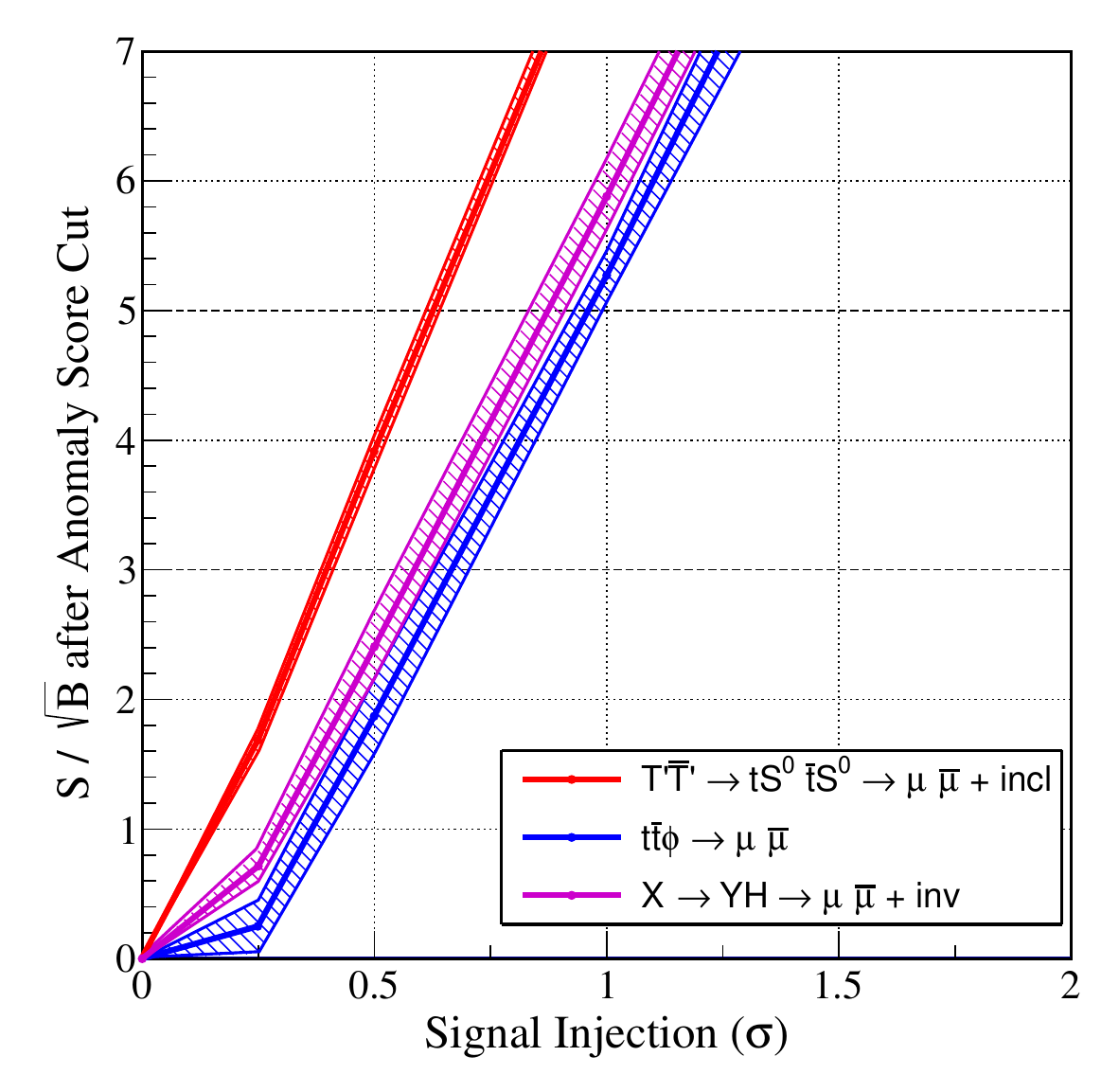}
    \includegraphics[width=0.44\linewidth]{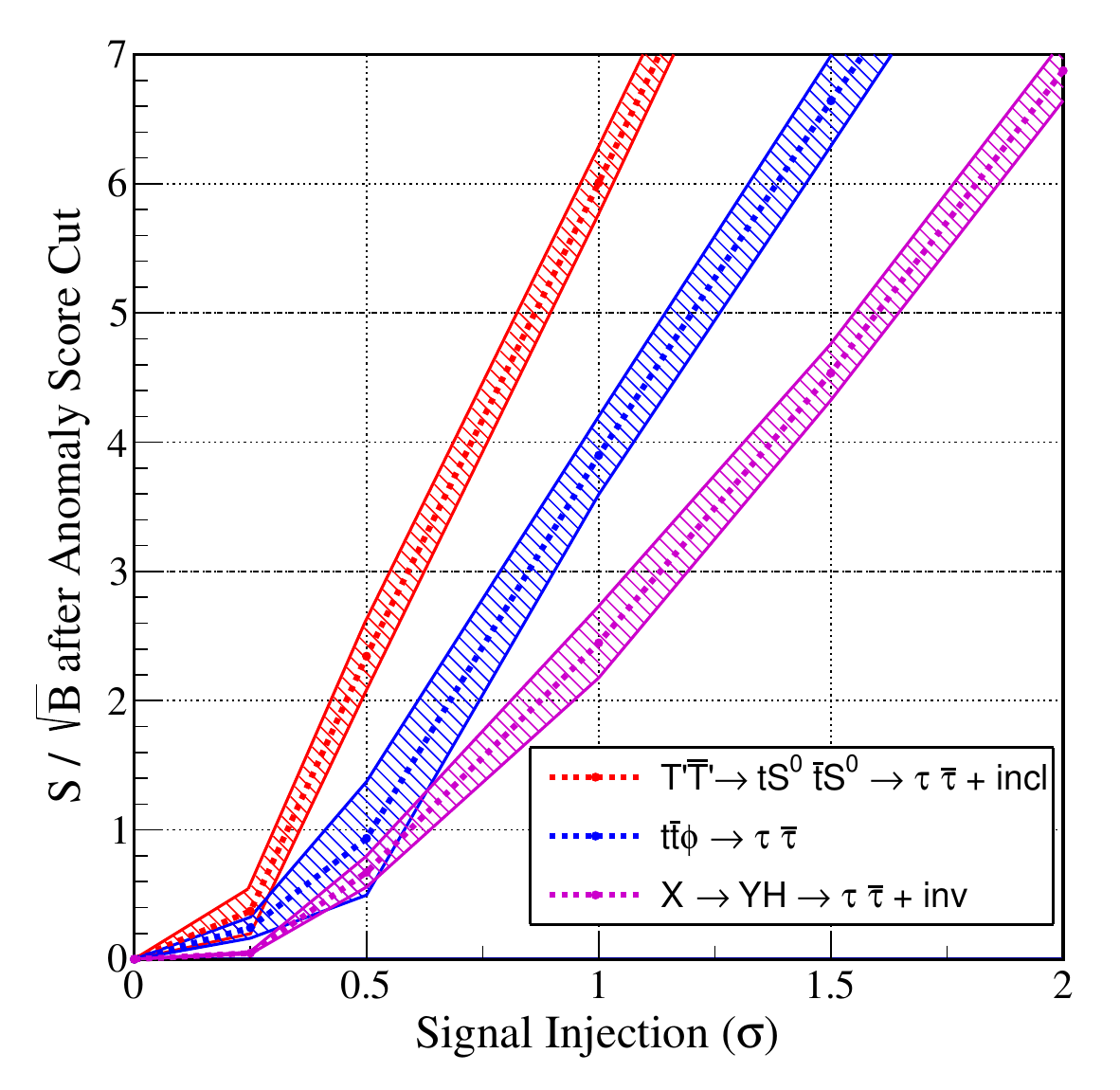}
    \caption{
    A summary of the sensitivity enhancement of our anomaly detection approach for both di-$\mu$ (left) and di-$\tau$ (right).
    The x-axis shows the size of the signal, quantified as $\frac{S}{\sqrt{B}}$ prior to any anomaly detection selection.
    The y-axis shows the approximate signal significance achieved after the application of an anomaly score cut that is 1\% efficient on the background.
    Our procedure is able to enhance the signal significance above the discovery threshold ($5 \sigma$) for moderately sized signal injections $\leq 2 \sigma$ for all considered signal models.}
    \label{fig:inj_summary}
\end{figure*}

Figure~\ref{fig:inj_summary} shows a summary of the anomaly detection performance for the different signal injections.
The x-axis denotes the strength of the signal injection and the y-axis the approximate signal significance after the anomaly detection procedure has been performed. 
The final significance of a signal is estimated as the significance improvement of the anomaly score cut, chosen to have a 1\% background efficiency, times the original injection strength.
For all three signal models considered, our procedure is able to enhance the signal significance above the discovery threshold ($5\sigma$) for moderately sized signal injections $\leq 2\sigma$. 
This demonstrates the potential of our anomaly detection approach to discover signals that may have been below the sensitivity threshold of a standard inclusive bump hunt search.

\section{Conclusion}
We have demonstrated the potential for weakly supervised anomaly detection to be applied to a new class of resonance searches.
Di-object plus~X searches, in which one looks for a resonance decaying to two standard model particles, but produced in association with other anomalous event activity, represents a new class of searches ripe for the application of anomaly detection.
In this exploratory work, we focused on di-$\tau$ and di-$\mu$ plus~X final states. 
We have employed new physically-motivated phase space variables to robustly capture salient features of multi-particle kinematics. 
We demonstrated that our anomaly detection approach can reach discovery-level significances for signals that would be missed in a conventional bump-hunt approach. 
Further work can be done to improve the generative modeling of topologically interesting variables like the hypersphere phase space coordinates. 
We could potentially gain sensitivity from exploring other event-level variables that quantify object type to complement the kinematic information of the phase space variables.
It may be fruitful to explore non-Euclidean normalizing flows, which may better capture the topological features of the phase-space variables.

The framework introduced in this paper lays the foundation for a program of searches that can be performed at the LHC with significant discovery potential.
Further exploration of topologies suitable for weakly supervised anomaly detection will be essential to deliver on the potential of anomaly detection to transform collider search strategies.  

\section*{Acknowledgments}

We would like to thank Joe Incandela, Danyi Zhang and Sanjit Masanam for useful discussions. We use the \verb|SK_CATHODE|~\cite{sk_cathode} library for implementation. We used the UCSB computational facilities administered by the Center for Scientific Computing at the California NanoSystems Institute and Materials Research Laboratory (an NSF MRSEC; DMR-1720256) and purchased through NSF CNS-1725797. UCSB is supported by the US Department of Energy under grant DE-SC0011702. Support for UCSB is also made possible by the Joe and Pat Yzurdiaga endowed chair in experimental science. LB was also supported by the National Science Foundation Graduate Research Fellowship Program under Grant No. (NSF 2139319).
OA is supported by FermiForward Discovery Group, LLC under Contract No. 89243024CSC000002 with the U.S. Department of Energy, Office of Science, Office of High Energy Physics.
LM and MS acknowledge support by the Deutsche Forschungsgemeinschaft (DFG, German Research Foundation) under Germany’s Excellence Strategy – EXC2121 “Quantum Universe” – 390833306 as well as by the Bundesministerium für Bildung und Forschung under project 05H24GUA. SS and PM are supported by NSF Grant PHY2310072. TC gratefully acknowledges the support of the SCale Family Fund during the career transition period.  

\bibliography{main}

\clearpage
\appendix
\section{Implementation Details}
\label{app:details}
\subsection{Preprocessing}
\label{subsec:preprocess}

The \verb|Delphes| files generated in Section~\ref{sec:samples} are further processed before being utilized with \cathode.  A luminosity under sampling is applied so the effective luminosity of the background samples is scaled to match the 138~fb$^{-1}$ collected data in LHC Run-2. A random selection of events with yield as detailed in the upper section of Table~\ref{tab:LuminositySamples} are used for analysis. After under sampling, the two simulated background datasets are combined and preprocessed following the cuts discussed in Section ~\ref{sec:Methods}, with the resulting event distributions shown in the lower section of Table ~\ref{tab:LuminositySamples}.

\begin{table}[!ht]
    \centering
    \renewcommand{\arraystretch}{0.75}
     \setlength{\tabcolsep}{6pt} 
    \scalebox{0.8}{
    \begin{tabular}{ccc}
    \toprule[1.5pt]  
    Name & \#Events & Lumi-scaled \#events  \\
    \midrule[\heavyrulewidth]
    Drell--Yan & 3M & 2,397,067  \\ 
    \cmidrule{1-3}  
    $t\bar{t}$ & 2M & 1,861,869 \\ 
    \midrule[\heavyrulewidth]
        & Combined Lumi-scaled & After cuts\\
    \cmidrule{1-3} 
    $\tau \bar{\tau}$ Drell--Yan + $t\bar{t}$ & 4,258,936 & $\sim$158,000  \\ 
    \cmidrule{1-3}  
    $\mu \bar{\mu}$ Drell--Yan + $t\bar{t}$ & 4,258,936 & $\sim$1,325,000  \\ 
    \bottomrule[1.5pt]  
    \end{tabular}
    }
    \caption{Number of events before and after luminosity under sampling (top). Event counts before and after preprocessing cuts (bottom).}
    \label{tab:LuminositySamples}
\end{table}

In the \cathode\ method, the synthetic background samples generated by the normalizing flow should be indistinguishable from the true background events in the signal region. 
If this is not the case, then the weakly supervised classifier will fail to learn to identify the signal, instead learning to distinguish the real and synthetic background events. 

We employ a sequence of transformations to optimize the feature space for our generative model. 
First, all features are shifted and scaled to the range $x \in (0,1)$
Then a logit transformation ($\text{logit}(x)=\ln\left(\frac{x}{1-x}\right)$) is applied to convert bounded domains into unbounded ones.
The transformed features are then standardized by subtracting the mean and dividing by the standard deviation of the training set, respective to each feature.
This preprocessing pipeline ensures that the normalizing flow operates on a well-conditioned feature space while preserving the essential structure of the phase space manifold.

Simulated and synthetic data in the signal and sideband regions are further split into training/validation/testing datasets following the methods described in Section~\ref{sec:Methods}. The corresponding event counts can be seen in Table~\ref{tab:SR_SB_split}.

\begin{table}[!ht]
    \centering
    \setlength{\tabcolsep}{6pt} 
    \scalebox{0.75}{\begin{tabular}{cccc}
    \toprule[1.5pt]  
    Name & Total \# of Events & Sideband & Signal Region  \\
    \midrule[\heavyrulewidth]
    $\tau\bar{\tau}$ Drell--Yan + $t\bar{t}$ & $\sim$158,000 & $\sim$127,000 & $\sim$31,000 \\ 
    \cmidrule{1-4} 
    $\mu\bar{\mu}$ Drell--Yan + $t\bar{t}$ & $\sim$1,325,000 & $\sim$1,260,000 & $\sim$65,000 \\
    \midrule[1.2pt]  
    $\tau\bar{\tau}$ & Training & Validation & Testing \\
    \midrule
    Sideband & 89,000 & 38,000 & - \\
    \cmidrule{1-4} 
    Signal Region & \begin{tabular}{@{}c@{}}13,000 Simulated\\39,000 Synthetic\end{tabular} & 
    \begin{tabular}{@{}c@{}}13,000 Simulated\\39,000 Synthetic\end{tabular} & 
    \begin{tabular}{@{}c@{}}5,000 Background\\5,000 Signal\end{tabular} \\
    \midrule[1.2pt]  
    $\mu\bar{\mu}$ & Training & Validation & Testing \\
    \midrule
    Sideband & 882,000 & 378,000 & - \\
    \cmidrule{1-4} 
    Signal Region & \begin{tabular}{@{}c@{}}27,500 Simulated\\82,500 Synthetic\end{tabular} & 
    \begin{tabular}{@{}c@{}}27,500 Simulated\\82,500 Synthetic\end{tabular} & 
    \begin{tabular}{@{}c@{}}10,000 Background\\10,000 Signal\end{tabular} \\
    \bottomrule[1.5pt]  
    \end{tabular}
    }
    \caption{Numbers of events used for training, validation, and testing in the signal region and sidebands (rounded to the nearest thousand).}
    \label{tab:SR_SB_split}
\end{table}

\subsection{Generative Model}
\label{subsec:gen_model}

Generative models with conditional flow matching (CFM) use conditional normalizing flows by learning vector fields between conditional distributions. CFM avoids the computational intractability of traditional flow matching objectives~\cite{lipman2023flowmatchinggenerativemodeling} while offering improved efficiency and stability over diffusion models~\cite{Krause:2024avx, lipman2023flowmatchinggenerativemodeling}.

Our flow is parameterized by a multi-layer perceptron with 6 hidden layers of 128 neurons each, using ELU activation functions~\cite{clevert2016fastaccuratedeepnetwork}.
Time is encoded using sinusoidal embeddings with 3 frequencies, producing a 6-dimensional time representation through concatenated sine and cosine components.
The network takes as an input the concatenation of the features $\vec{x}$, time embeddings $t$ and conditioning variable $m$ (di-$\tau$ or di-$\mu$ visible invariant mass).

Training employs the Adam optimizer~\cite{kingma2017adam} with a cosine annealing learning rate scheduler, starting at $1 \times 10^{-3}$ and decaying to $1 \times 10^{-6}$.
We used batch sizes of 1024 (di-$\mu$) and 512 (di-$\tau$).
We trained the model over 2,000 to 5,000 epochs using A100 GPUs. We found that training beyond 5,000 epochs did not improve the quality of the generated samples. We fine tuned the hyperparameters of the generative model by minimizing the ROC between background and generated samples. Training for 5,000 epochs takes 32 hours.

\begin{figure}[!ht]
    \includegraphics[width=0.98\linewidth]{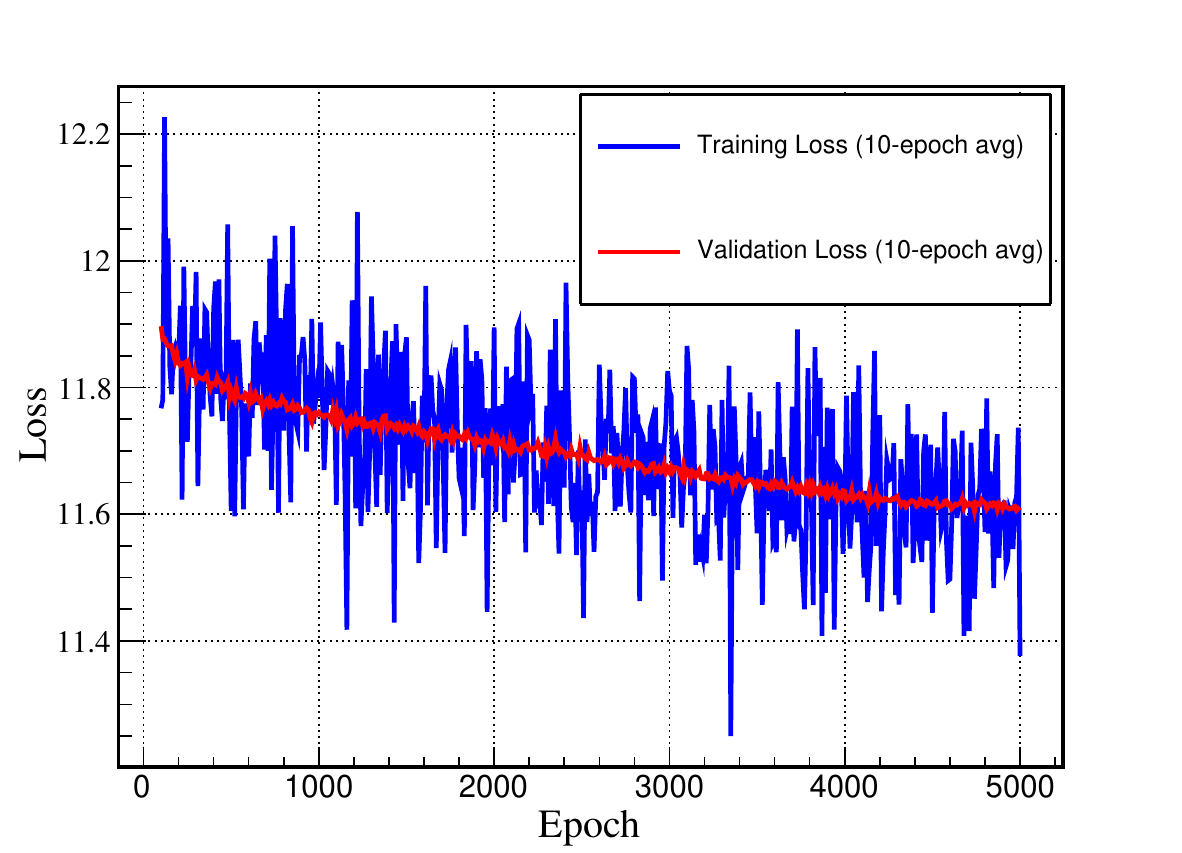}
    \caption{Training and validation loss for the di-$\mu$ channel, as an example.}
    \label{fig:training_and_validation_loss}
\end{figure}

After training, we select the model state with the lowest validation loss for the subsequent interpolation and sampling steps. The training and validation losses of the di-$\mu$ background can be seen in Fig.~\ref{fig:training_and_validation_loss}.

To generate synthetic background events in the signal region, we employ a two-step process that requires the trained flow to extrapolate beyond its training domain: 
(1) We fit a kernel density estimate to the di-$\ell$ invariant mass distribution observed in the signal region in  the training set using a Gaussian kernel with bandwidth 0.01. 
(2)~We sample mass values $m$ from this signal region mass distribution and use them to condition the learned normalizing flow that was trained on sideband data.
This approach requires the learned normalizing flow to extrapolate from sideband masses (training domain) to signal region masses (target domain), assuming the conditional dependence on $m$ generalizes smoothly across this mass range.

Since the normalizing flow operates on transformed features, we apply the inverse of the standardization and logit transformation to convert the sampled events back to physical space.

We do not model the hypersphere coordinates in this analysis due to their non-Euclidean nature, where normalizing flows are designed for features in the d-dimensional Euclidean space $\mathcal{R}^d$ ~\cite{lipman2024flowmatchingguidecode}.
The flow struggled to learn how to model the connection between different coordinates needed for a good background distribution.
The simplex still captures much of the event's topology, but the additional azimuthal information contained in the hypersphere could improve identification.
There are limited works looking into flows on non-Euclidean manifolds~\cite{liu2023,falorsi2021, rezende2020}. 
However, implementing and improving upon their architectures for 10-dimensional hyperspheres is beyond the scope of this paper.

\section{Comparison to $p_T$-based features}
\label{sec:pt_analysis}

To evaluate the effectiveness of the phase space representation, we conducted a comparative analysis replacing the 10 simplex coordinates used with the $p_T$ values of the 10 highest $p_T$ objects in each event, in descending order. 
We focus on the di-$\mu$ channel as a proof of concept for this comparison.

To maintain consistency with the phase space analysis, events with fewer than 10 objects were zero-padded using randomized exponentially decreasing values from 0 to 25\,GeV, corresponding to the minimum $p_T$ threshold in \verb|Delphes| simulation.
We retain the same complementary kinematic features used in the phase space analysis: the scalar sum of jet $p_T$, the missing transverse momentum (MET), and the angular separation $\Delta R$ between the two muons. 
These features provide essential event-level information not captured by the ranked $p_T$ values alone.

\begin{figure*}
    \centering
    \includegraphics[width=0.32\linewidth]{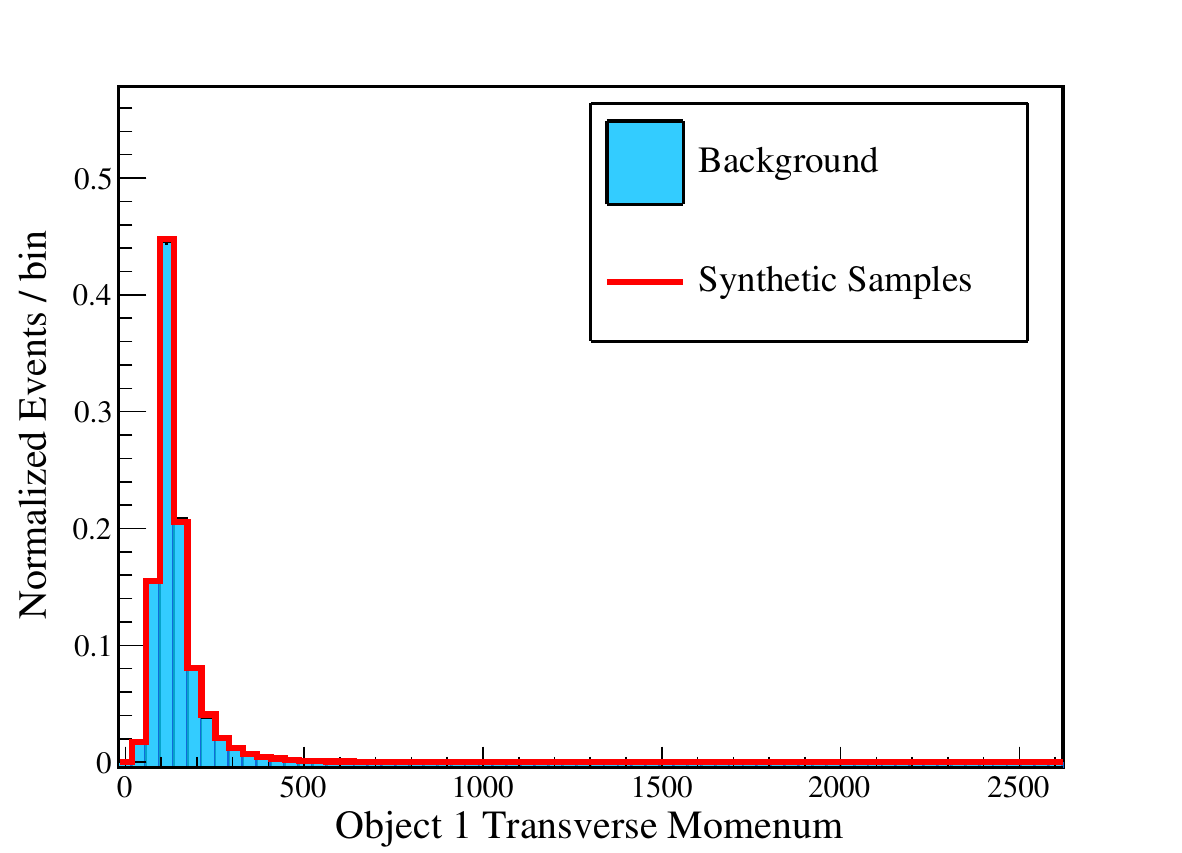}
    \includegraphics[width=0.32\linewidth]{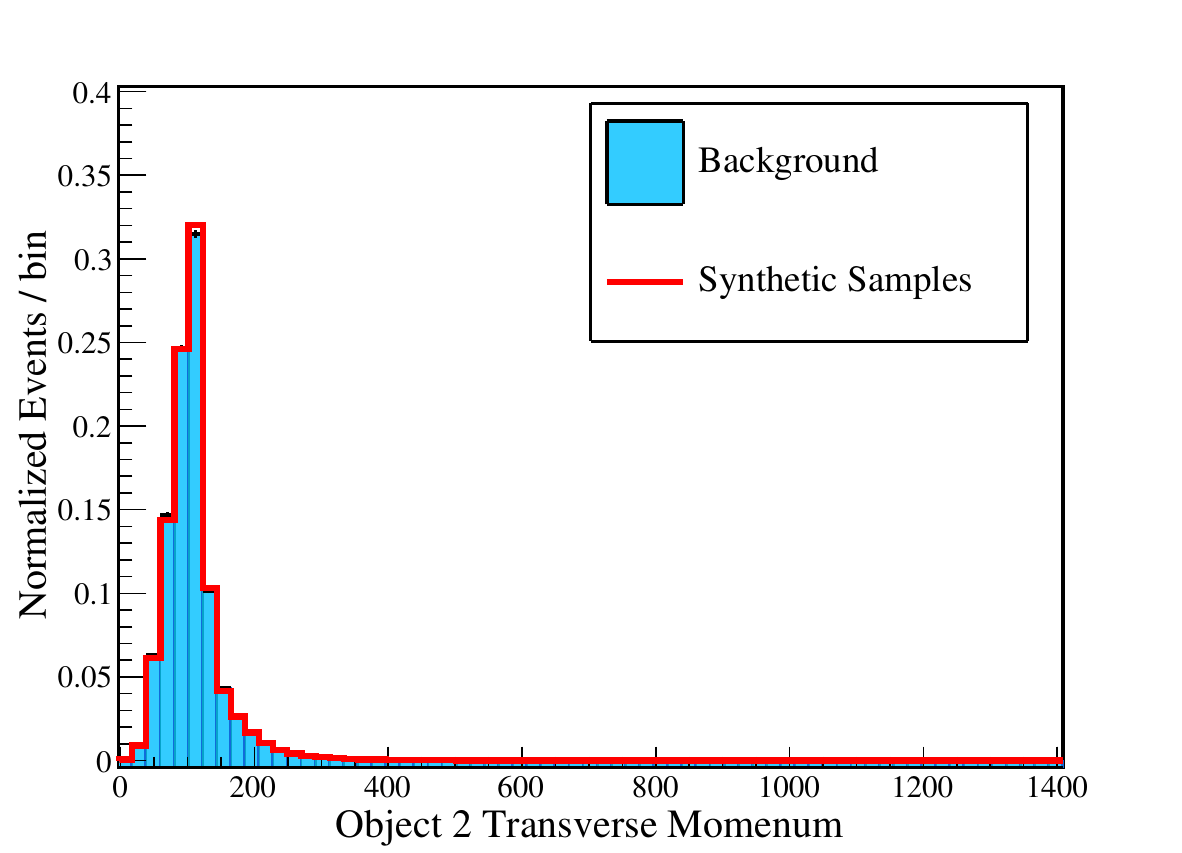}
    \includegraphics[width=0.32\linewidth]{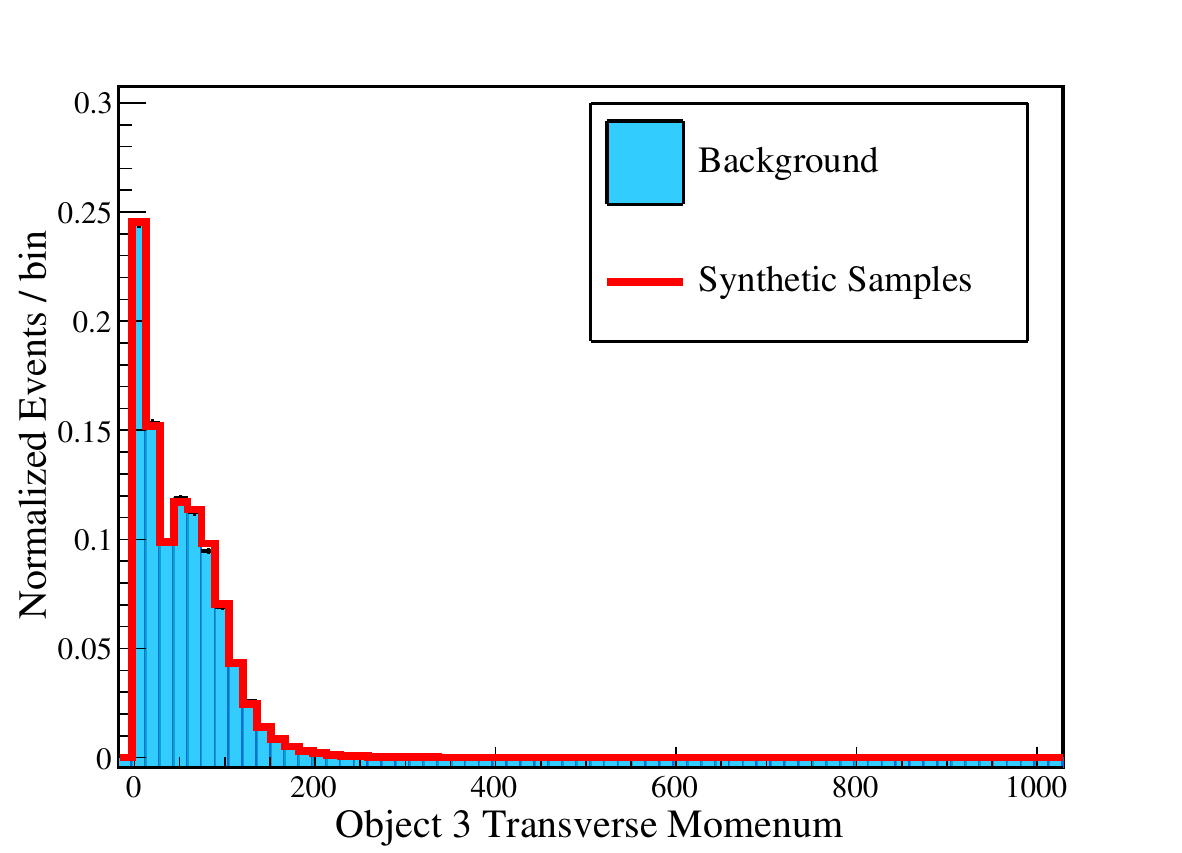} \\
    \includegraphics[width=0.32\linewidth]{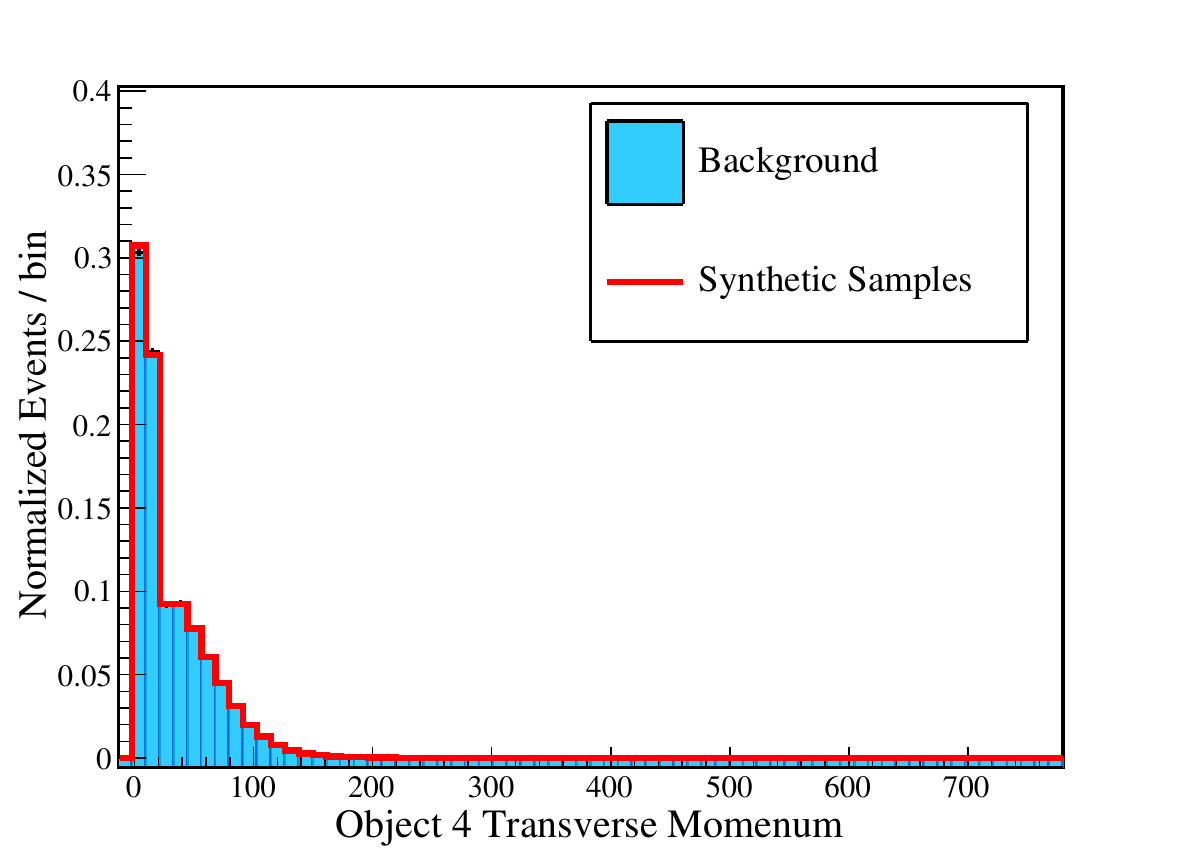}
    \includegraphics[width=0.32\linewidth]{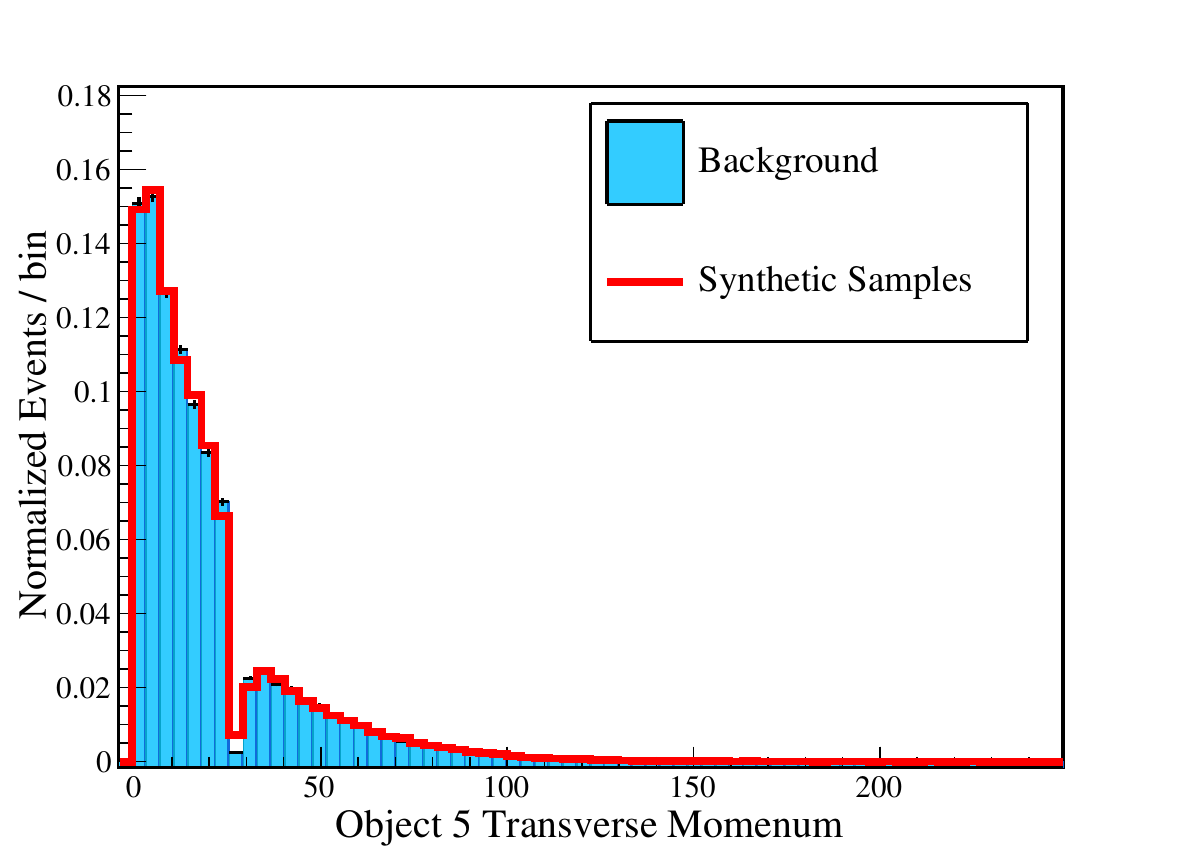}
    \includegraphics[width=0.32\linewidth]{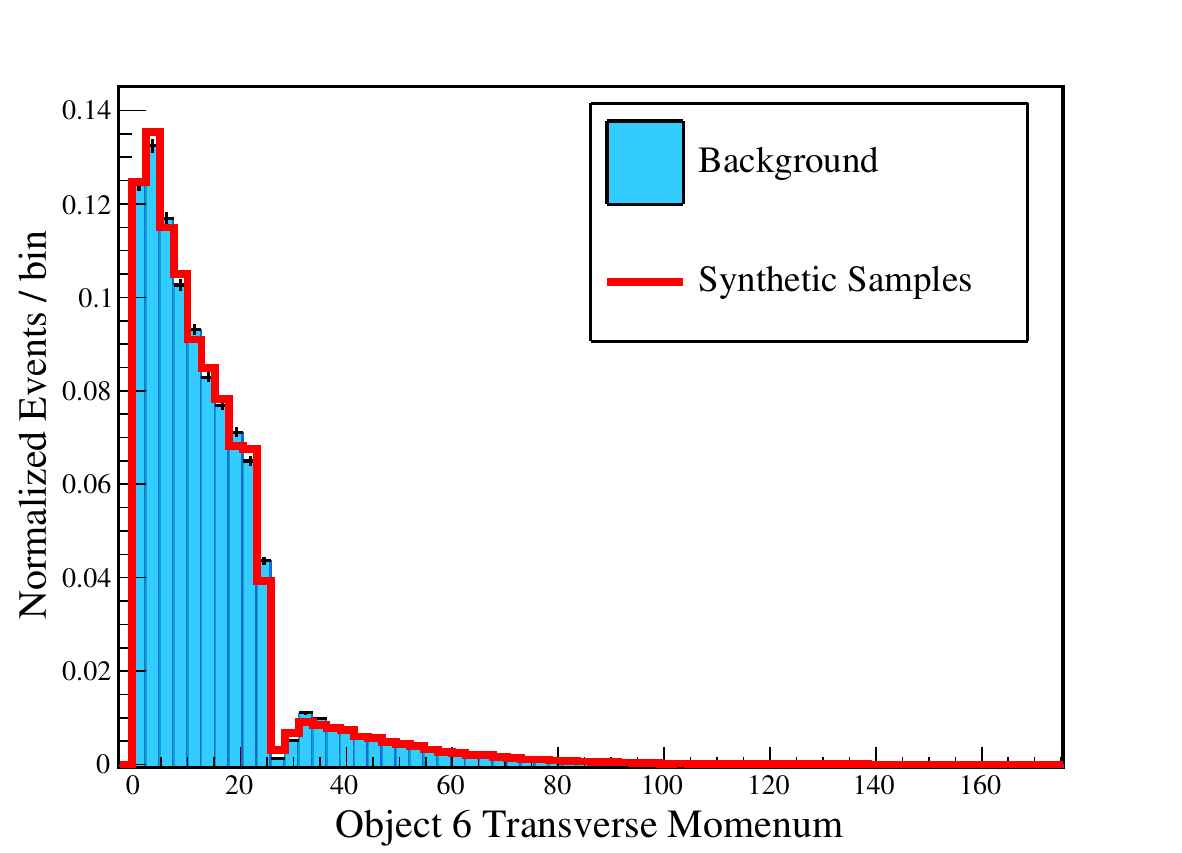} \\
    \includegraphics[width=0.32\linewidth]{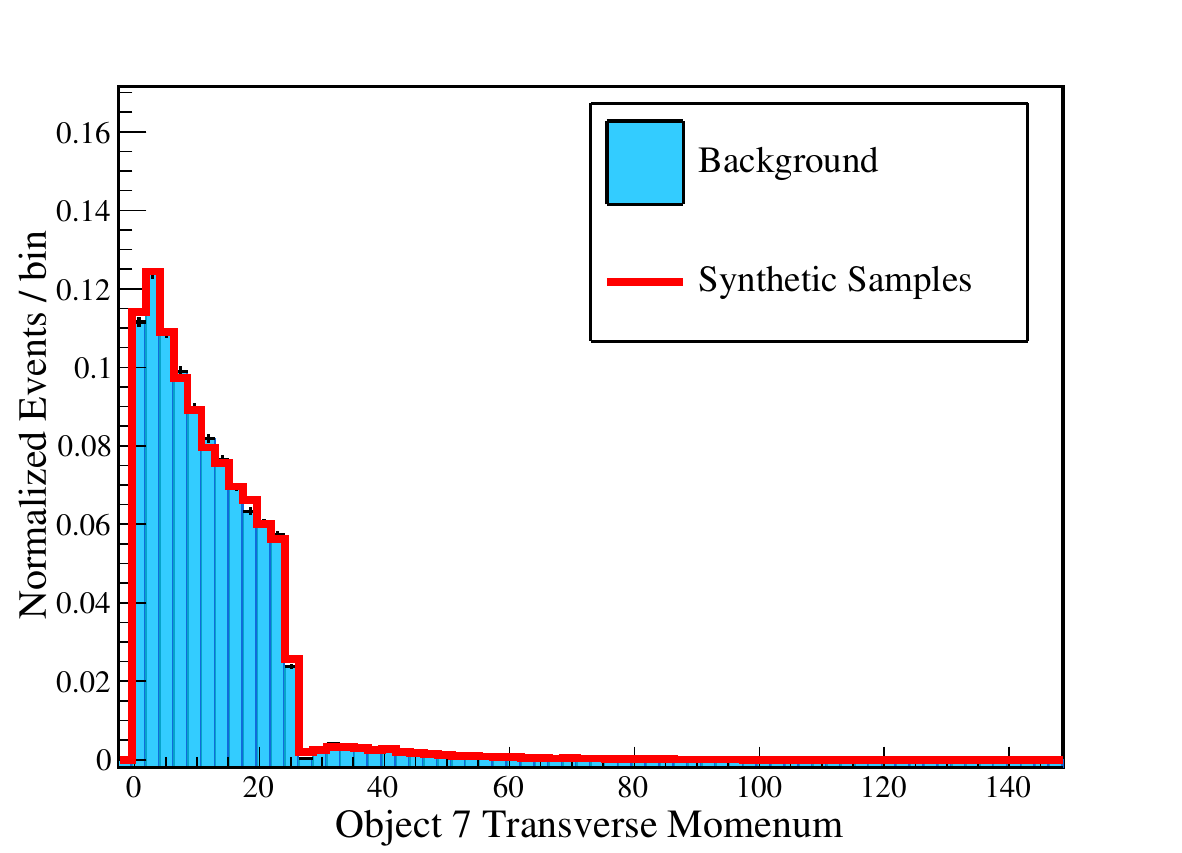}
    \includegraphics[width=0.32\linewidth]{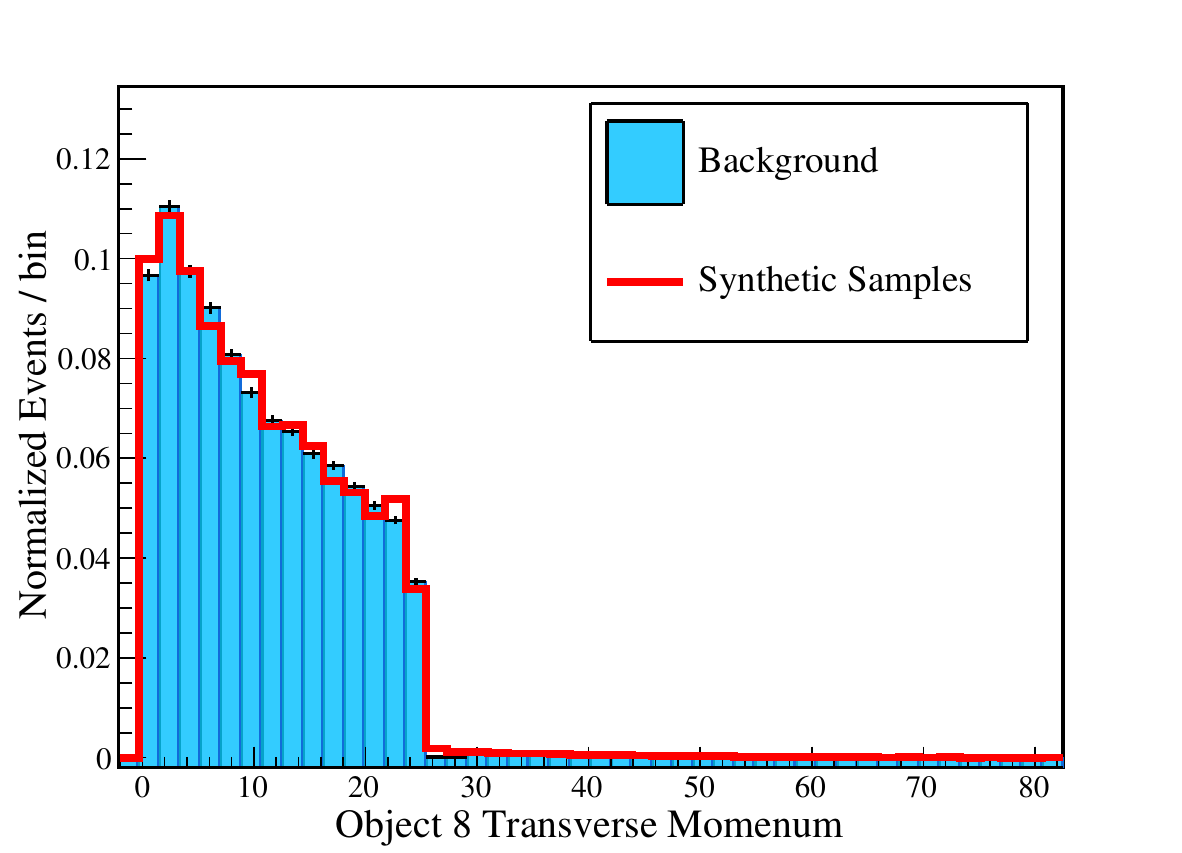}
    \includegraphics[width=0.32\linewidth]{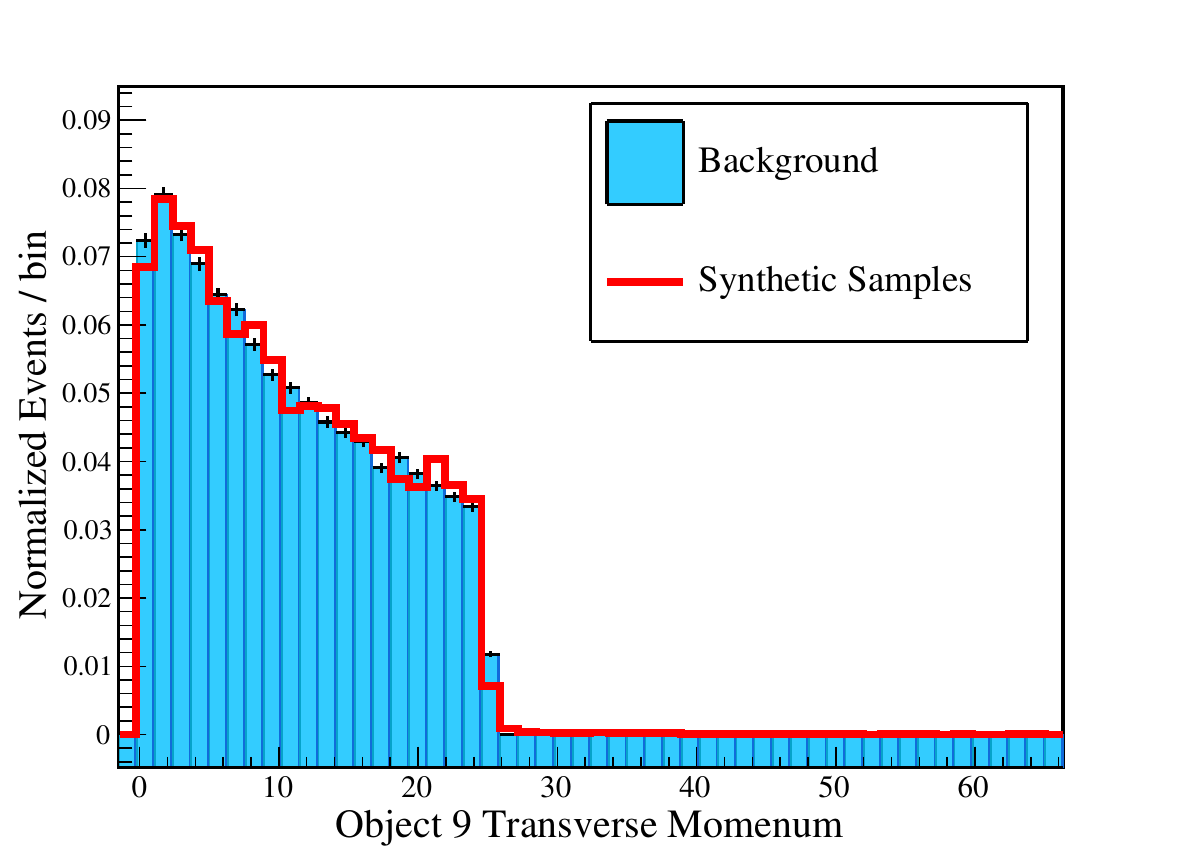} 
    \includegraphics[width=0.32\linewidth]{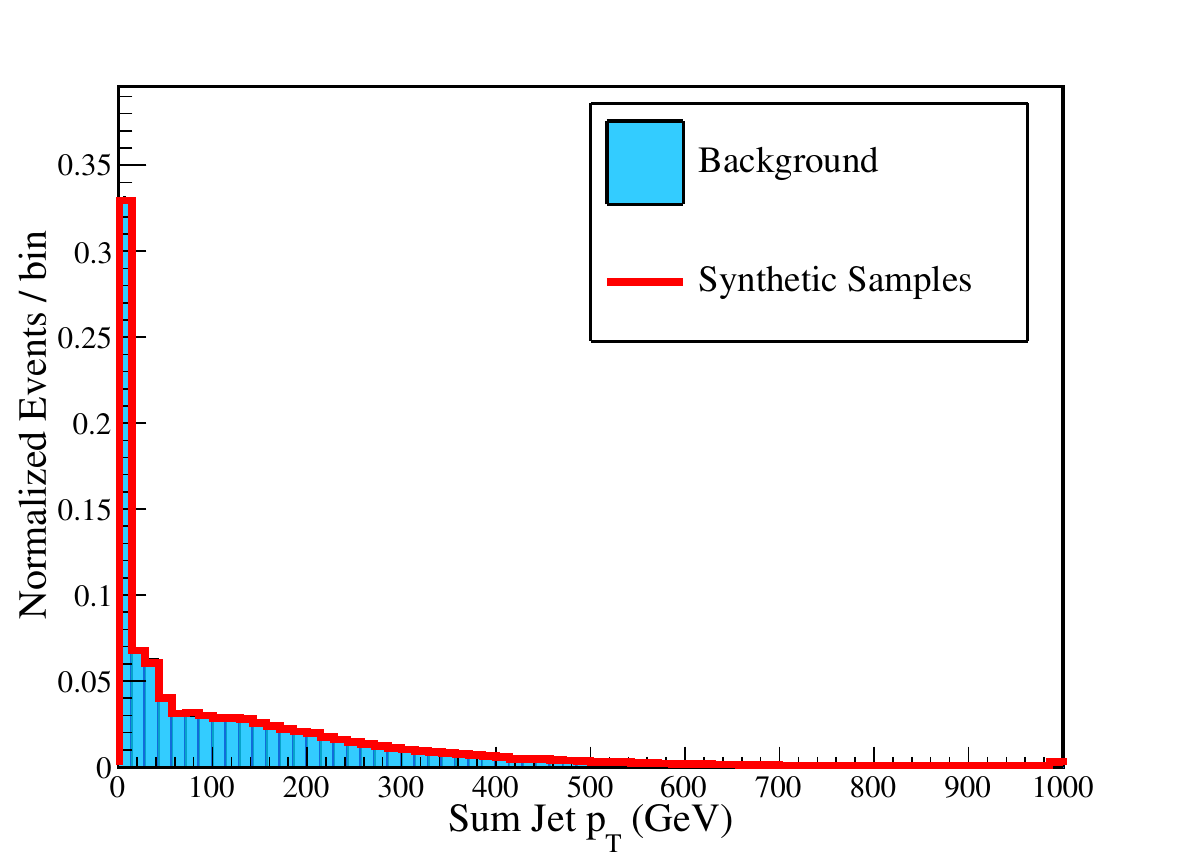}
    \includegraphics[width=0.32\linewidth]{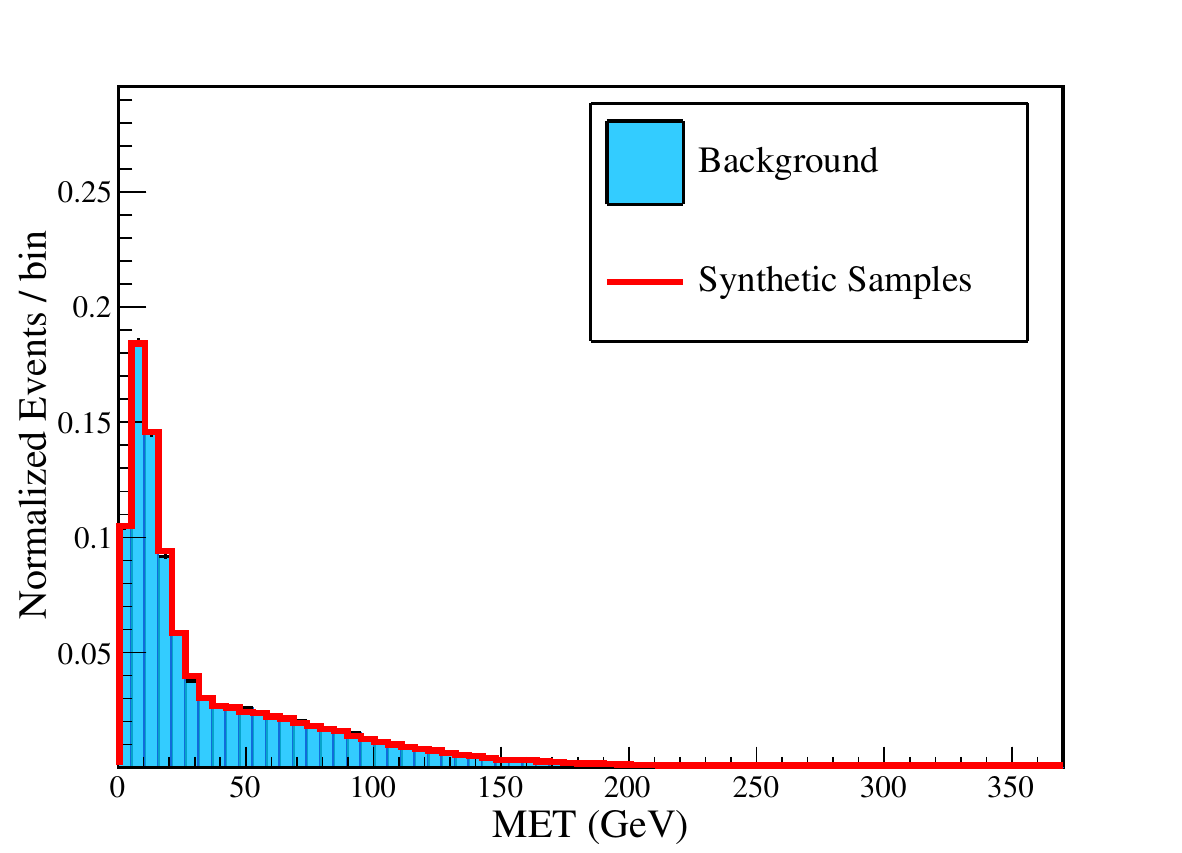} 
    \includegraphics[width=0.32\linewidth]{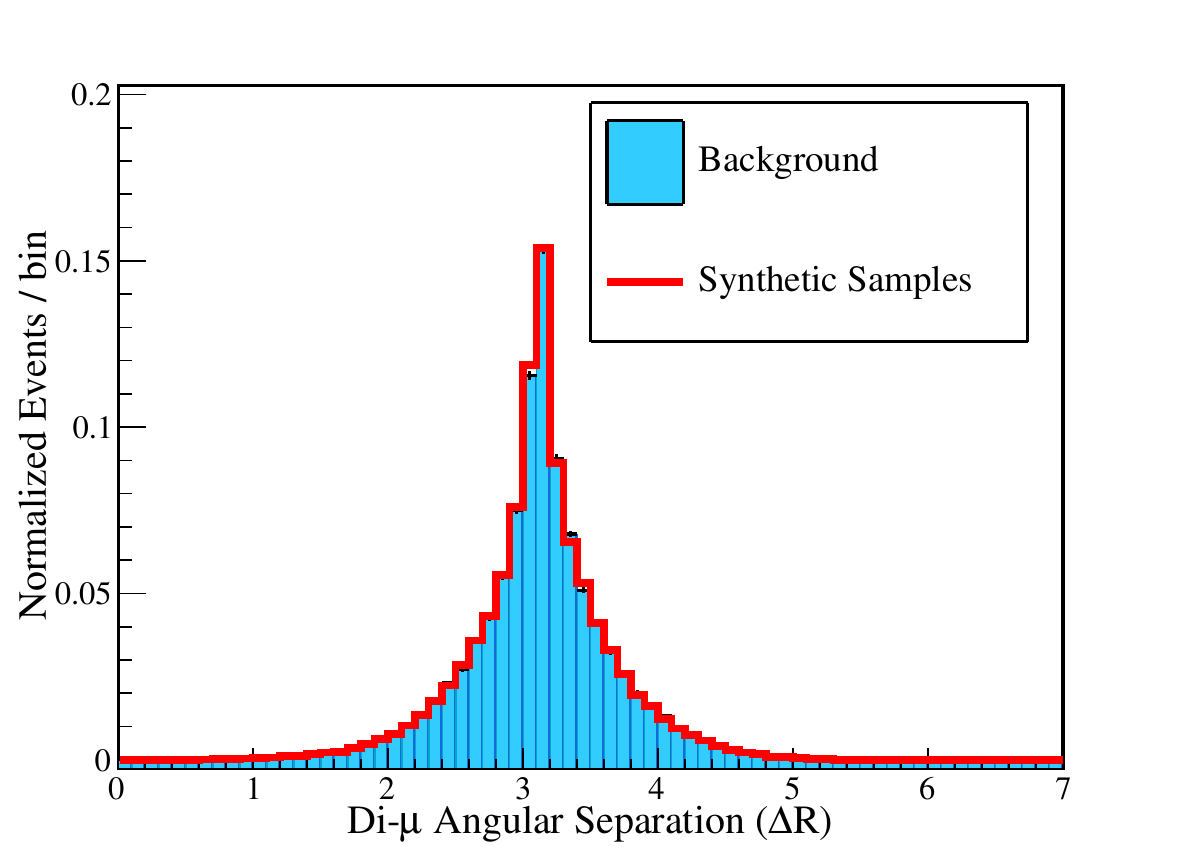} 

    \caption{A comparison of the distributions of true background events in the signal region and the synthetic ones generated by our normalizing flow for di-$\mu$. Object 10 $p_T$ is omitted for brevity, but has similar distributions.}
    \label{fig:p_T_feats}
\end{figure*}

The features can be seen in Figure~\ref{fig:p_T_feats}.

The same generative model architecture and training methodology employed for the phase space di-$\mu$ analysis were applied to the $p_T$-based features. 
The background generation achieved a ROC score of 0.53, comparable to the phase space background model performance (ROC = 0.52), indicating similar quality in modeling the background distribution.

\begin{figure*}[!ht]
    \includegraphics[width=0.7\linewidth]{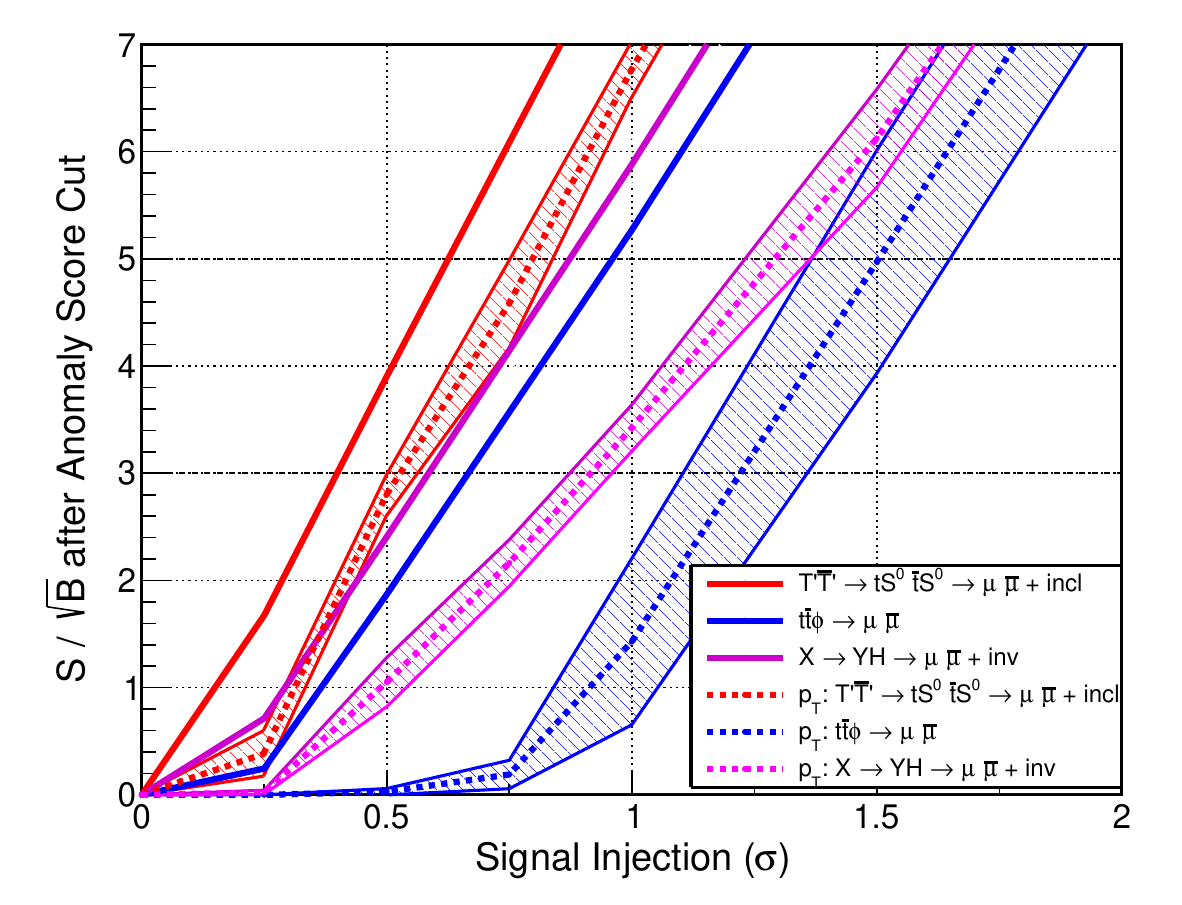}
    \caption{A summary of the sensitivity enhancement of our anomaly detection approach for di-$\mu$ using phase space features (solid line) and $p_T$ features (dashed line).
    The x-axis shows the size of the signal, quantified as $\frac{S}{\sqrt{B}}$ prior to any anomaly detection selection.
    The y-axis shows the approximate signal significance achieved after the application of an anomaly score cut that is 1\% efficient on the background.
    Our procedure utilizing phase space features is superior to using the $p_T$ of the objects alone in all signal cases.}
    \label{fig:p_T_Results}
\end{figure*}

Figure~\ref{fig:p_T_Results} shows the resulting significance improvement curves comparing the $p_T$ and phase space approaches. 
The phase space coordinates consistently outperform the $p_T$-based features across all signal injection levels.

The superior performance of phase space coordinates is encouraging for several reasons. 
First, the simplex coordinates are specifically designed to capture the topological structure of multi-particle final states in a coordinate-invariant manner, whereas $p_T$ rankings only preserve energy ordering without topological information.
Second, our analysis utilizes only the simplex coordinates from the phase space decomposition and we have not yet incorporated the hypersphere coordinates, which encode additional angular correlations. 
This suggests significant untapped potential for further improvements in anomaly detection sensitivity using the complete phase space representation.
The consistent performance advantage of phase space coordinates across different signal models demonstrates the robustness of this approach and supports its adoption for anomaly detection in collider physics applications.

\end{document}